%% file: compReview.tex
\documentclass[final,5p,times,twocolumn]{elsarticle}
\usepackage{amssymb}
\usepackage{xspace}
\usepackage{units}
\usepackage{xcolor}
\usepackage{subfigure}
\usepackage{booktabs}
\usepackage{amsmath,amssymb}
\usepackage[colorlinks=false]{hyperref}
\hypersetup{
colorlinks,%
citecolor=blue,%
filecolor=blue,%
linkcolor=blue,%
urlcolor=blue
}

\journal{Astroparticle Physics}

\include{abbreviations}

\biboptions{sort&compress}

\begin{document}

\begin{frontmatter}

%%%%%%%%%%%%%%%%%%%%%%%%%%%%%%%%%%%%%%%%%%%%%%%%%%
\title{Measurements of the Cosmic Ray
Composition with Air Shower Experiments}

\author[khk]{Karl-Heinz Kampert}
\author[mu]{Michael Unger}

\address[khk]{Bergische Universit\"at Wuppertal, Fachbereich C / Physik, Gaussstr. 20, D-42119 Wuppertal}
\address[mu]{Karlsruher Institut f\"ur Technologie, Institut f\"ur Kernphysik, Postfach 3640
76021 Karlsruhe}
%%%%%%%%%%%%%%%%%%%%%%%%%%%%%%%%%%%%%%%%%%%%%%%%%%

\begin{abstract}
In this paper we review air shower data related to the
mass composition of cosmic rays above \energy{15}. After explaining the
basic relations between air shower observables and the primary mass
and energy of cosmic rays, we present different approaches and results of composition
studies with surface detectors. Furthermore, we discuss measurements
of the longitudinal development of air showers from non-imaging Cherenkov detectors
and fluorescence telescopes.

The interpretation of these experimental results in terms of primary
mass is highly susceptible to the theoretical uncertainties of
hadronic interactions in air showers. We nevertheless attempt to
calculate the logarithmic mass from the data using different hadronic
interaction models and to study its energy dependence from \energy{15}
to \energy{20}.
\end{abstract}


\end{frontmatter}

%% uncomment for line-numbering
% \linenumbers

%=================================================
%%%% Introduction
%=================================================
\section{Introduction}
\label{sec:introduction}
\input{introduction}

%=================================================
\section{Air Shower Observables Sensitive to Composition}
\label{sec:observables}
\input{observables}

%=================================================
\section{Measurements of the Nuclear Composition}
\label{sec:results}
Strictly speaking, no air shower observatory measures the primary
composition of cosmic rays. Instead, one or more of the mass sensitive
observables from the last section can be measured and the data
can then be {\itshape interpreted} in terms of primary mass by a comparison
to air shower simulations using hadronic interaction models.
Since different air shower observables react differently to changes
in the characteristics of hadronic interactions, one may hope to diminish
the model dependence of primary mass estimates by comparing the results
from different observables.

In this section we will therefore first review measurements of
composition related air shower observables from particle detectors at
ground and from optical detectors. The data will then be compared to
each other in terms of the average logarithmic mass in
Sec.~\ref{sec:lnA}.
\subsection{Particle Detectors}
\label{sec:particleDet}
\input{particleDet}

\subsection{Non-imaging Cherenkov Detectors}
\label{sec:nicDets}
\input{nicDets}

\subsection{Fluorescence Telescopes}
\label{sec:fluoDets}
\input{fluoDets}

%-------------------------------------------------------
% ===  summary of all measurements in terms of <lnA> ===
%-------------------------------------------------------
\subsection{Average Logarithmic Mass}
\label{sec:lnA}
\input{lnA}

%=================================================
%%%% Photons and Neutrinos
%=================================================
\section{Search for Neutral Primaries}
\label{sec:neutral}
Measurements of neutral primaries, i.e.\ neutrons, photons, and
neutrinos provide additional crucial information about the
acceleration models and sources of cosmic rays as well as on
their propagation through the universe. Unlike charged cosmic rays they are not deflected by magnetic fields and could point back to their sources. Specifically, high energy neutrinos pointing back to a source would provide the first unambiguous signal of a high energy cosmic ray accelerator. Moreover, a `guaranteed' flux of high energy photons and neutrinos is provided by the GZK-effect (see Sec.~\ref{sec:introduction}). An observation of such cosmogenic neutrinos or photons is considered as a `smoking gun' that would complement the observation of the flux suppression seen in the cosmic ray energy spectrum.
Neutrons arise at all energies from charge exchange interactions at the source or in the interstellar medium. However, their lifetime limits their propagation distance to about $10\,\mathrm{kpc}\cdot (E/\!\EeV)$.

Separating neutron- from proton-induced showers is not possible. Nevertheless, samples of air showers can be enriched with respect to light primaries and searches for neutron point sources can be performed. Such studies have yielded no detection yet and are reviewed in \cite{Sommers-11}. Separating photon- or neutrino-induced EAS from nuclei-induced showers is experimentally much easier than distinguishing light from heavy nuclear primaries.

In this chapter we will review recent photon and neutrino searches aimed at highlighting their complementarity to measurements of the nuclear composition.

\input{neutral}

%=================================================
%%%% Conclusions
%=================================================
\section{Conclusions}
\label{sec:conclusions}
\input{conclusions}

%=================================================
%%%% acknowledgement
%=================================================
\section{Acknowledgement}
We would like to thank S.P.\ Knurenko and V.\ Prosin for providing
data tables and valuable information on the Yakutsk and Tunka data and
the Pierre Auger and TA Collaboration for permission to use their data
prior to journal publication.  Furthermore we would like to express
our gratitude to our colleagues in the Pierre Auger and KASCADE-Grande
collaborations for many inspiring discussions about the cosmic ray
composition and we appreciate the critical reading of the manuscript of
this article by R.\ Clay, C.\ Dobrigkeit, D.\ Veberi\v{c} and the
anonymous journal referee.

\bibliographystyle{elsarticle-num}
\bibliography{compReview}

\end{document}

%% file: abbreviations.tex
\def\MeV{\ifmmode {\mathrm{\ Me\kern -0.1em V}}\else
                   \textrm{Me\kern -0.1em V}\fi\xspace}%
\def\GeV{\ifmmode {\mathrm{\ Ge\kern -0.1em V}}\else
                   \textrm{Ge\kern -0.1em V}\fi\xspace}%
\def\TeV{\ifmmode {\mathrm{\ Te\kern -0.1em V}}\else
                   \textrm{Te\kern -0.1em V}\fi\xspace}%
\def\PeV{\ifmmode {\mathrm{\ Pe\kern -0.1em V}}\else
                   \textrm{Pe\kern -0.1em V}\fi\xspace}%
\def\EeV{\ifmmode {\mathrm{\ Ee\kern -0.1em V}}\else
                   \textrm{Ee\kern -0.1em V}\fi\xspace}%
\def\eV{\ifmmode {\mathrm{\ e\kern -0.1em V}}\else
                   \textrm{e\kern -0.1em V}\fi\xspace}%
\def\gcm{\ifmmode {\mathrm{g/cm}^2}\else
                   {g/cm$^2$}\fi\xspace}%
\def\Xmax{\ifmmode {X_\mathrm{max}}\else
                   {$X_\mathrm{max}$}\fi\xspace}%
\def\sigmaXmax{\ifmmode {\sigma(X_\mathrm{max})}\else
                   {$\sigma(X_\mathrm{max})$}\fi\xspace}%
\def\meanXmax{\ifmmode {\langle X_\mathrm{max}\rangle}\else
                   {$\langle X_\mathrm{max}\rangle$}\fi\xspace}%
\def\meanLnA{\ifmmode {\langle\ln A\rangle}\else
                   {$\langle\ln A\rangle$}\fi\xspace}%
\newcommand\meanXmaxX[1]{\langle X_\mathrm{max}^{#1}\rangle}

\newcommand{\EcritEM}{\varepsilon_\mathrm{c}^\mathrm{em}}
\newcommand{\EcritPi}{\varepsilon_\mathrm{d}^\pi}
\newcommand{\energy}[1]{\unit[$10^{#1}$]{\eV}}

%% file: introduction.tex
Knowledge of the mass composition of cosmic rays is of key importance
for solving the long standing puzzle about the origin of high-energy
cosmic rays. The mass (and therefore charge) distribution can provide
strong constraints on the acceleration of cosmic rays and on their
propagation through the galactic and extragalactic Universe. Of
particular interest are measurements of the mass composition in
vicinity of structures observed in the energy spectrum of cosmic rays,
most prominently the ``knee'' and the ``ankle'' at approximately
\energy{15.5} and \energy{18.5}, respectively, as well as at energies
above \energy{19.5} at which a flux suppression --- possibly the
GZK-effect --- has been observed.  Correlated changes of the energy
spectrum and mass composition can provide important clues to the
origin of these features. For instance, a rigidity dependent
cut-off~\cite{peters61} of the spectra of cosmic rays with different
charge can lead to a gradual increase in the average mass of galactic
(see~\cite{Hoerandel:2004gv} and references therein) and
extra-galactic cosmic rays~\cite{Allard:2008gj, Aloisio:2009sj,
Hooper:2009fd}.  On the other hand, the interpretation of the ankle
and flux suppression as a signature of propagation
effects~\cite{greisen-66, Zatsepin:1966jv} of ultra-high energy
extragalactic protons would require a very light composition above the
ankle~\cite{Berezinsky:2002nc}.  Furthermore, the location and nature
of the transition from galactic to extragalactic cosmic rays (or lack
thereof~\cite{Calvez:2010uh}) leads to distinct predictions of the
energy evolution of the mass composition of cosmic rays in the energy
region between \energy{17} and \energy{19} (see
e.g.~\cite{Wibig:2004ye,Allard:2005cx,Hillas:2005cs,
Aloisio:2007rc,DeDonato:2008wq,Berezhko:2009mf}).

Up to energies of some \energy{14}, the cosmic ray composition can be
measured directly and with only minor, or none disturbing effects from
cosmic ray interactions in the atmosphere, by employing stratospheric
balloon- or satellite-borne experiments, respectively.  Cosmic rays at
these energies are covered elsewhere, see e.g.~\cite{castellina-00}
or~\cite{Seo-11} and~\cite{Boezio-11} in this volume.  At higher
energies, measurements of cosmic rays are only possible via
observations of extensive air showers (EAS). Amongst all EAS
measurements, those aiming at reconstructing the cosmic ray mass
composition are the most difficult ones. This is, because the mass of
the primary particle can only be inferred from detailed comparisons of
experimental observables with air shower simulations, with the latter
being subject to uncertainties of hadronic interactions at the highest
energies. Fortunately, the data from the LHC at CERN will allow for
detailed tests of interaction models up to center of mass energies
being equivalent to cosmic ray energies of \energy{17}, but an
appreciable amount of extrapolation of the properties of hadronic
interactions will still be needed to interpret the air shower data up
to energies of \energy{20}. In fact, these model uncertainties become
a dominant source of systematics of estimates of the cosmic ray mass
composition at the highest energies. Therefore, identifying different
EAS observables with sensitivity to the primary mass is of great
importance as these different observables should lead to consistent
conclusions about the properties of the primary
particles. Inconsistent conclusions, on the contrary, would be
indicative for deficiencies of the employed hadronic interaction
models.

High energy photons and neutrinos can provide an indirect approach to
the composition of cosmic rays at the highest energies. This is
because photo-pion production via the Greisen-Zatsepin-Kuzmin
effect~\cite{greisen-66, Zatsepin:1966jv} leads to emission of
neutrinos~\cite{Berezinsky-69} and photons from pion decays in the
reactions $p+\gamma_\mathrm{CMB} \to p + \pi^0 \to p + \gamma\gamma$,
or likewise $p+\gamma_\mathrm{CMB} \to n+\pi^+ \to n+\mu^+ + \nu_\mu
\to n+e^+ + \nu_e + \bar{\nu}_\mu + \nu_\mu $.  The discovery of
cosmogenic photons and neutrinos would prove the existence of the GZK
energy loss mechanism independently from the cut-off in the cosmic ray
energy spectrum and would be the unambiguous consequence of a light
cosmic ray composition at ultra-high energies. In the case of heavy
primaries, the photo-pion production sets in at much higher energies,
because the energy threshold for this reaction depends on the energy
per nucleon and not the total energy.  Instead, photo-disintegration
is the dominant process~\cite{puget-76} for energy losses, giving rise
to neutrinos from neutron decays, but yielding much lower neutrino
fluxes in the EeV range compared to the photo-pion production by
primary protons.

This paper is organized as follows: In the next section we will
provide an overview of the most important air shower observables with
respect to measurements of the nuclear composition. In
Sec.~\ref{sec:results} experimental results covering the energy range
from the knee to ultra-high energies will be presented and searches
for high energy photons and neutrinos will be briefly discussed in
Sec.~\ref{sec:neutral}.  In this review we focus on measurements of
the cosmic ray composition from air shower experiments only. Other
aspects of cosmic rays above the knee are discussed elsewhere in this
volume~\cite{Sommers-11,Aloisio-11, Allard-11} and
in~\cite{Nagano:2000ve, Bluemer-09, Kotera:2011wo, Letessier-11b}.

%% file: observables.tex
There exist a plethora of experimental techniques to characterize air
shower features. In the context of mass composition at least
two orthogonal measurements are needed to estimate both, the energy
and mass, of the primary cosmic ray that initiated the EAS. This is
usually achieved by observing either the longitudinal development of a
shower or by the simultaneous determination of the electromagnetic
and muonic component of EAS at ground level. Further insights
can be gained by the study of the lateral distribution of particles
at ground
level, which is related to the longitudinal development
stage of the shower at observation level.

The ability to deduce the nuclear composition of cosmic rays on a
statistical basis from these measurements relies to a large extent on
the theoretical understanding of the shower development and the
hadronic interactions that occur within the cascade.  However, the
major differences between air showers induced by different primary
masses have a rather simple cause.  The discriminating power
originates from the fact that a primary nucleus of mass $A$ and energy
$E$ can in good approximation be treated as a superposition of $A$
nucleons of energy $E^\prime = E/A$.  This {\itshape superposition
  model} seems plausible because the binding energy of nucleons is
much smaller than the energy of the primary particle. However, the
nucleons of a primary nucleus are obviously not independent. For
instance, the first interaction of a nucleus of mass $A$ will happen
on average at a depth $\langle X_\mathrm{first} \rangle$ equal to the
interaction length $\lambda_A<\lambda_p$ and the average
number of nucleons participating in the first interaction, $\langle
n_A \rangle$, fluctuates.  Interestingly enough, it can be
shown~\cite{semisuper} that for $\langle n_A \rangle = A\,
\lambda_A/\lambda_p$ (the average number of participating
nucleons from Glauber theory~\cite{Glauber:1970jm}), the inclusive
distribution of first interactions follows exactly the expectation of
the naive superposition model, namely that the average depth of
interaction of the $A$ nucleons is $\lambda_p$. This is the
{\itshape semi-superposition theorem} that provides a justification of
the superposition ansatz under more realistic assumptions.

Whereas the superposition model can explain
 the main differences between air showers induced by
different nuclei, it does of course not account for nuclear effects
such as re-interaction in the target nucleus or for the effect of nuclear
fragmentation on air shower fluctuations (see
e.g.~\cite{Kalmykov:1989br, Kalmykov:1993qe, Schatz:1994hv}).
More realistic predictions of air showers can be obtained by using
transport codes like {\scshape Corsika}~\cite{soft:CORSIKA}, {\scshape
  Aires}~\cite{Sciutto:1999jh} or {\scshape Cosmos}~\cite{Roh:2011st}
together with hadronic interactions models such as {\scshape
  Epos}~\cite{Pierog:2006qv}, {\scshape
  QGSJet}~\cite{Ostapchenko:2010vb} or {\scshape
  Sibyll}~\cite{Ahn:2009wx} (see~\cite{Knapp:2002vs} for a
comprehensive review of air shower simulations).

While these transport codes can give an accurate description of air
showers given a model for hadronic interactions, it is often desirable
to qualitatively understand the basic physics behind a certain air
shower observable. For this purpose it can be constructive to
calculate the air shower development within the
Heitler-model~\cite{CarlsonOppenheimer, Heitler} in which the cascade
is approximated by a simple deterministic branching model. The
original treatment of electromagnetic cascades has recently been
extended to hadronic showers in~\cite{AlvarezMuniz:2002ne,
  Matthews:2005sd, Horandel:2006jd} and some of these qualitative
dependencies will be referred to in the following.

\subsection{Longitudinal Development}
\label{sec:longDev}
The observation of the longitudinal development of the particle
cascade in the atmosphere is especially well suited for composition
studies.  In analogy to particle physics detectors, the atmosphere
acts as a huge homogeneous (as opposed to sampling) calorimeter in
which the air is both the passive material that drives the shower
development and the active material that allows the detection of the
produced particles via fluorescence or Cherenkov light
(cf. Sec.~\ref{sec:nicDets} and~\ref{sec:fluoDets}), as well as the light
guide through which the signal is transferred to the detector.

\begin{figure*}[t!]
  \centering
  \includegraphics[width=0.65\linewidth]{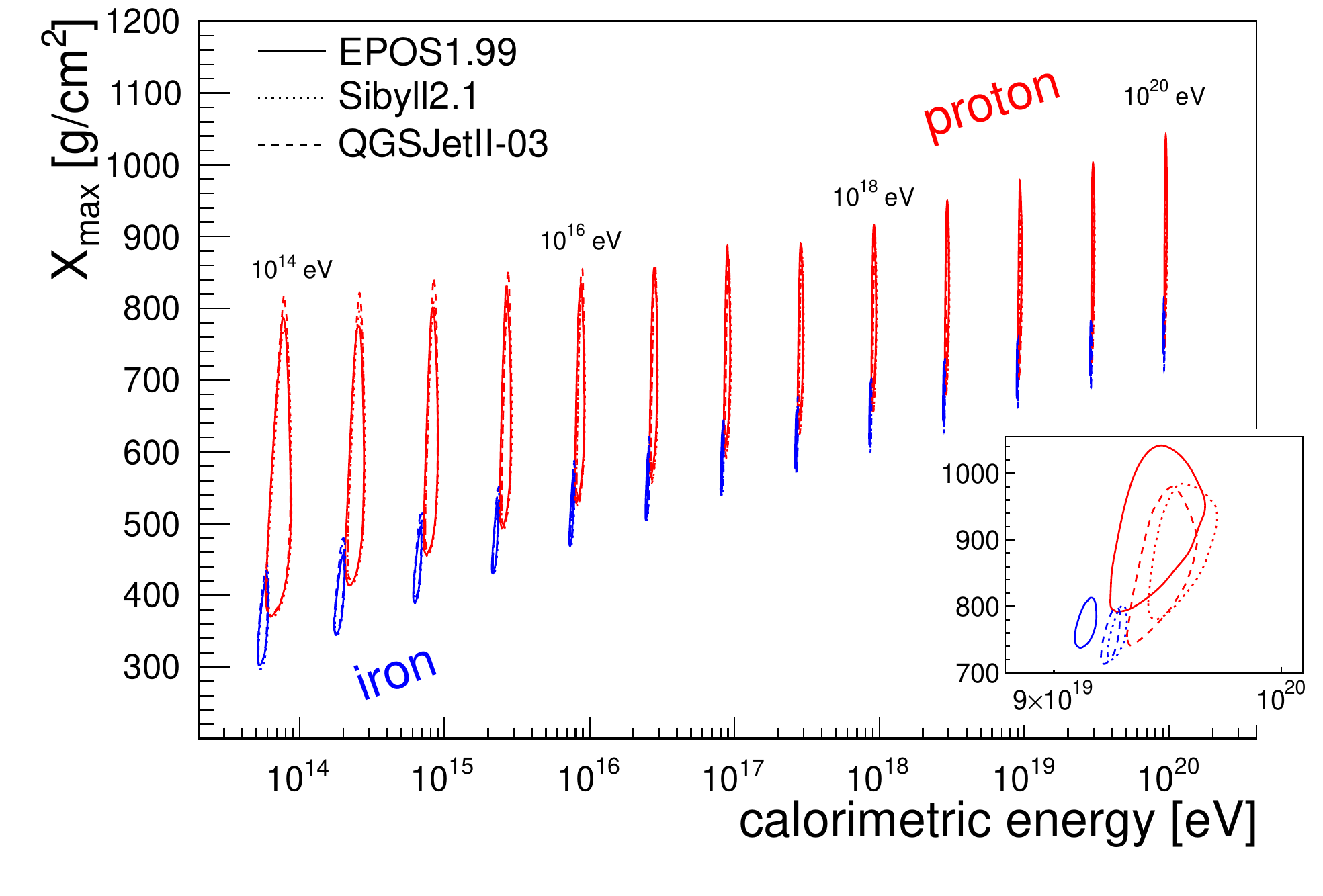}
  \caption[compoVariablesLong]{Air shower simulation
    of the shower maximum vs.\ calorimetric energy. Contour
    lines illustrate the regions which include 90\,\% of the showers and
    the inset shows a detailed view at \energy{20}.}
  \label{fig:compoVariablesLong}
\end{figure*}

The first inelastic interaction of the primary particle of energy $E$
and mass $A$ with an atmospheric nucleus occurs at an average depth
which is equal to interaction length for inelastic nucleus-air
collisions, $\lambda_{A-\mathrm{air}}$.  In this and each consecutive
interaction about one third of the primary energy is transferred from
the hadronic to the electromagnetic component of the shower via decays
of neutral pions into photons. The energy transfer continues until the
charged hadrons decouple from the shower by decay into muons and
neutrinos, i.e.\ until their interaction length becomes larger than
their decay length (see next section).  If the charged hadrons start
to decay on average after $n_\mathrm{d}$ interaction lengths, then the
energy in the electromagnetic component is
\begin{equation}
  E_\mathrm{cal} \approx E \left(1-\left(\frac{2}{3}\right)^{n_\mathrm{d}}\right).
\end{equation}
Since $n_\mathrm{d}\ge 5$ for energies
$\ge$\energy{15}~\cite{Meurer:2005dt}, it follows that most of the
energy of an air shower can be observed in its electromagnetic part
and it is this so-called {\itshape calorimetric energy} which allows
detectors that can observe the longitudinal air shower development to
estimate the primary energy with good accuracy.

Given this estimator of the energy of the primary particle, the
orthogonal variable sensitive to its primary mass is the slant depth
at which the particle cascade reaches its maximum in terms of the
number of particles, $\Xmax$. This {\itshape shower maximum} is
dominated by electromagnetic sub-showers produced in the interaction
with the largest inelasticity which is usually (though not always) the
first interaction. An electromagnetic shower of energy $E$ reaches its
maximum at about
\begin{equation}
  \langle X_\mathrm{max}^\mathrm{emag} \rangle \approx X_0 \ln{E/\EcritEM}
\end{equation}
where $X_0 \approx 36.62$\,\gcm is the radiation length in air and
$\EcritEM \approx 84$\,\MeV is the critical energy in air at which
ionization and bremsstrahlung energy losses are equal. If the total
multiplicity of hadrons produced in the main interaction is $N$ and
the average hadron energy is $E/N$, then the shower maximum of a
primary proton is
\begin{equation}
  \langle X_\mathrm{max}^p\rangle \approx
  \lambda_p + X_0 \ln\left(\frac{E}{2 N \EcritEM}\right)
\label{eq:protonXmax}
\end{equation}
where both the hadronic interaction length and particle production
multiplicity are energy dependent. The factor 2 takes into account
that neutral pions decay into two photons.  Furthermore, the shower
maximum is expected to be influenced by the {\itshape elasticity} of
the first interaction, $\kappa_\mathrm{ela} = E_\mathrm{lead}/E$,
where $E_\mathrm{lead}$ is the energy of the highest energy secondary
produced in the interaction.  For interactions with
$\kappa_\mathrm{ela}>0.5$ most of the primary energy will be
transferred deeper into the atmosphere and correspondingly the shower
maximum will be deeper. We are not aware of a consistent treatment of
the elasticity within a Heitler model for the longitudinal
development, however using air shower simulations, the dependence on
the elasticity fits well to
\begin{equation}
  \langle X_\mathrm{max}^p\rangle \approx
  \lambda_p + X_0 \ln\left(\frac{\kappa_\mathrm{ela} E}{2 N 
\EcritEM}\right).
\label{eq:protonXmax2}
\end{equation}

The {\itshape elongation rate}~\cite{elong1, elong2, elong3} is a
measure of the change of the shower maximum per logarithm of energy,
\begin{equation}
  D = \frac{\text{d}\langle X_\mathrm{max}\rangle}{\text{d}\ln E}.
\end{equation}
For protons and constant elasticity Eq.~(\ref{eq:protonXmax2}) gives
\begin{equation}
  D_\mathrm{p} = \frac{\mathrm{d} \langle 
  X_\mathrm{max}^p\rangle}{\mathrm{d}\ln E}
  \approx X_0\, (1-B_N - B_\lambda)
\end{equation}
where the changes in multiplicity and interaction length are given by
\begin{equation}
  B_N = \frac{\mathrm{d} \ln N}{\mathrm{d} \ln E} \quad \mathrm{and} \quad
  B_\lambda = - \frac{\lambda_p}{X_0} \frac{\mathrm{d} \ln 
\lambda_p}{\mathrm{d} \ln E}.
\label{eq:BFactors}
\end{equation}
Since hadronic interaction models predict an approximately logarithmic
decrease of $\lambda_p$ with energy and $N\varpropto
E^\delta$, $D_p$ is approximately constant and therefore
\begin{equation}
 \langle X_\mathrm{max}^p\rangle \approx c +  D_p \ln E,
\end{equation}
with parameters $c$ and $D_\mathrm{p}$ being dependent on the
characteristics of hadronic interactions.  Using the aforementioned
\mbox{(semi-)superposition} assumption, one obtains
\begin{eqnarray}
\langle X_\mathrm{max}^{A}\rangle= \langle 
X_\mathrm{max}^p(E^\prime=E/A)\rangle =
c +  D_p \ln(E/A)
\label{eq:xmaxA}
\end{eqnarray}
and at a given energy the average shower maximum for a mixed composition with fractions
$f_i$ of nuclei of mass $A_i$ is
\begin{equation}
  \langle X_\mathrm{max}\rangle
  \approx  \sum_i f_i \, \langle X_\mathrm{max}^{A_i}\rangle
  =  \langle X_\mathrm{max}^p\rangle - D_p \,\langle\ln A\rangle.
\label{eq:lnA}
\end{equation}
This equation explicitly demonstrates the relation of $\meanXmax$ to
the {\itshape average logarithmic mass} of the cosmic ray composition,
$\langle\ln A\rangle = \sum_i f_i \, \ln A$.

The numerical value of $D_p$ from air shower simulations is
about 25\,\gcm (or about 60\,\gcm for the change in $\meanXmax$ per
decade, $D_{10}^p = \ln(10)\, D_p$) and therefore
proton and iron induced air showers are expected to differ by around
$D_p (\ln 56 - \ln 1 ) \approx 100\,\gcm$. Moreover, if the
hadronic cross sections and multiplicities rise with energy (and if
there are no sudden changes in the elasticity as for instance
suggested in~\cite{AlvarezMuniz:2006ye}), then Eq.~(\ref{eq:BFactors})
leads to Linsley's {\itshape elongation rate theorem} which states
that the value of $D_p$ cannot exceed the radiation length in
air, $X_0$. Therefore, Eq.~(\ref{eq:lnA}) implies that if an
experiment measures an elongation rate of $D>X_0$, then a change in
the cosmic ray composition from light to heavy, $\text{d} \langle\ln
A\rangle/\text{d}\ln E > 0$, must be responsible for that larger
value.

Results of air shower simulations of \meanXmax and $E_\mathrm{cal}$
are shown in Fig.~\ref{fig:compoVariablesLong}. As can
be seen, the calorimetric energy is indeed a good proxy for the
primary energy and exhibits only small shower-to-shower
fluctuations. And, as expected from the relations sketched above, the
shower maximum penetrates deeper into the atmosphere
with the logarithm of the energy and is
shallower for heavy nuclei than for light ones.  The shower-to-shower
fluctuations in \Xmax are however large and even extreme compositions
like pure proton vs.\ pure iron have a considerable overlap in their
$\Xmax$-distributions.

However, these fluctuations carry interesting information about the
primary particle types and/or the 'mixedness' of the cosmic ray
composition, that can be experimentally exploited. For protons the
standard deviation $\sigma$ of the $\Xmax$-distribution, $\sigma^2 =
\langle (X_\mathrm{max}^p)^2\rangle - \langle
X_\mathrm{max}^p\rangle^2$, is given by the quadratic sum of
the fluctuation of the first interaction point and the fluctuations of
the shower development, which in case of the simple Heitler model
Eq.~(\ref{eq:protonXmax2}) reduces to
\begin{equation}
\sigma^2_p \approx \lambda_p^2  +
                           \left( X_0 \frac{\sigma(N)}{N}\right)^2 +
                           \left( X_0 \frac{\sigma(\kappa_\mathrm{ela})}
                                  {\kappa_\mathrm{ela}}\right)^2.
\end{equation}
Interestingly, for a geometrical treatment of nucleus-nucleus
scattering~\cite{Bialas:1976ed}, the relative fluctuations of the
multiplicity, $\sigma(N)/N$, are constant. Furthermore, if the
relative fluctuations of the elasticity do not depend on energy, then
$\sigma^2_p$ is expected to decrease with energy reflecting
the logarithmic rise of the proton-air cross section.

For primary nuclei, the naive expectation from the superposition model
would be $\sigma_A = \sigma_p/\sqrt{N}$. As will be
seen in Fig.~\ref{fig:rmsplot}, air shower simulations predict a
$\sigma_p$ of about 60\,\gcm at \energy{18} and
correspondingly $\sigma_\mathrm{Fe}$ should be at the level of 8\,\gcm
whereas a much larger value of $\approx 20$\,\gcm is predicted. The
reason for this discrepancy is that although according to the
semi-superposition theorem the positions of elementary nucleon-nucleon
interactions are distributed $\varpropto \exp(-X/\lambda_p)$, the
individual positions are not independent.  For instance, the $n$
nucleons participating in the first interaction have all the same
$X=X_\mathrm{first}$. These correlations together with the
event-by-event variance of $n$ are the main reason for
$\sigma_A > \sigma_p/\sqrt{N}$. In addition there
are two extreme cases for the spectator nucleons: Either complete
fragmentation (i.e. all $A-n$ spectators propagate independently after
$X_\mathrm{first}$) or a single spectator nucleus with $A^\prime = A
-n$, where the latter case without fragmentation would result to an
additional broadening of the $\Xmax$-distribution.

In the interesting case of a mixed composition with fractions $f_i$ of
nuclei of mass $A_i$, the combined $\Xmax$-distribution follows
from elementary statistics~\cite{LinsleyXmax1983}. If each component
has an average shower maximum of $\meanXmax_i$ with width $\sigma_i$
then
\begin{equation}
  \sigma(\Xmax)^2 = \langle \sigma_i^2 \rangle +
        \left(\left\langle\meanXmax_{i}^2\right\rangle- \meanXmax^2\right)
\label{eq:sigmaMixed}
\end{equation}
\begin{figure}
  \includegraphics[width=\linewidth]{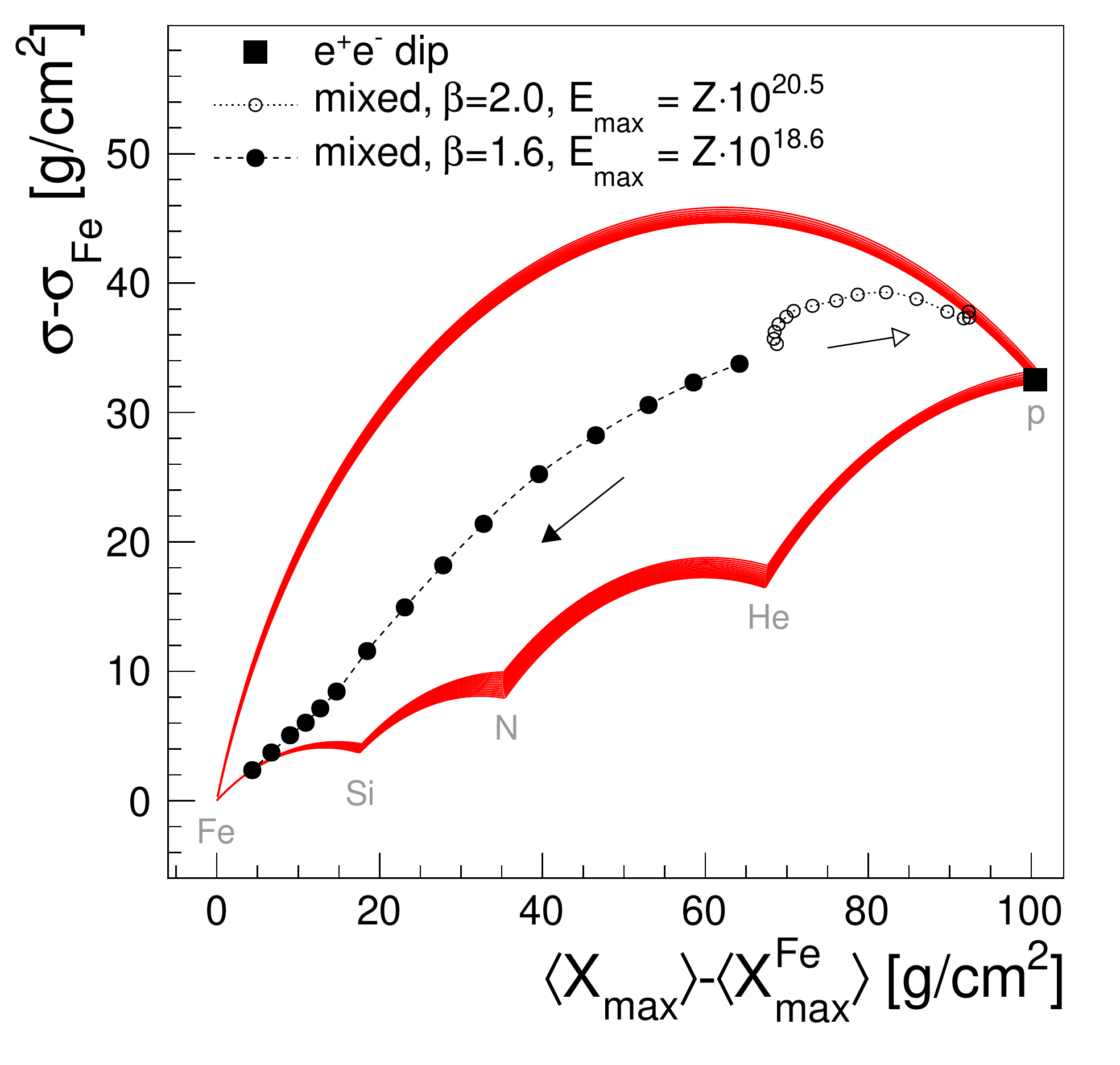}
  \caption[xmaxVsSigma]{Composition sensitivity of a combined
    measurement of the shower maximum and its fluctuations. Dots
    denote different models of the extragalactic cosmic ray
    composition at Earth: Pure proton composition in the
    dip-model~\cite{Berezinsky:2002nc} and mixed composition with a
    large~\cite{Allard:2005cx} or small~\cite{Allard-11} maximum
    energy of the sources and different spectral indices $\beta$ at
    the source. Energies range from \energy{18.5} to \energy{20} with
    a spacing of $\Delta\lg E=0.1$. Red lines are simulations with
    {\scshape Sibyll} at the same energies for various two-component
    transitions. The contour defined by these transition contains all
    other possible mixtures for $1\le A \le 56$.}
  \label{fig:sigmaVsXmax}
\end{figure}
where $\meanXmax = \left\langle\meanXmax_{i}\right\rangle$ is the mean
of the combined distribution. In case of a composition comprised of two components
only, this reduces to
\begin{equation}
  \sigma(\Xmax)^2 = f \sigma_1^2 + (1-f) \, \sigma_2^2 + f (1-f) (\Delta\meanXmax)^2.
\label{eq:sigmaTwoComp}
\end{equation}
 Therefore,
depending on the separation of the mean values $\Delta\meanXmax$ and
the fraction $f$ of component $1$, it can happen that the combined
distribution is {\itshape broader} than the individual distributions
because their separation adds to the total width.  In terms of average
logarithmic mass, Eq.~(\ref{eq:sigmaMixed}) can be rewritten as
\begin{equation}
  \sigma(\Xmax)^2 = \langle \sigma_i^2 \rangle +
  D_p^2 \left(\langle\ln^2 A\rangle- \meanLnA^2\right),
\end{equation}
\begin{figure*}[ht!]
  \centering
  \includegraphics[width=0.65\linewidth]{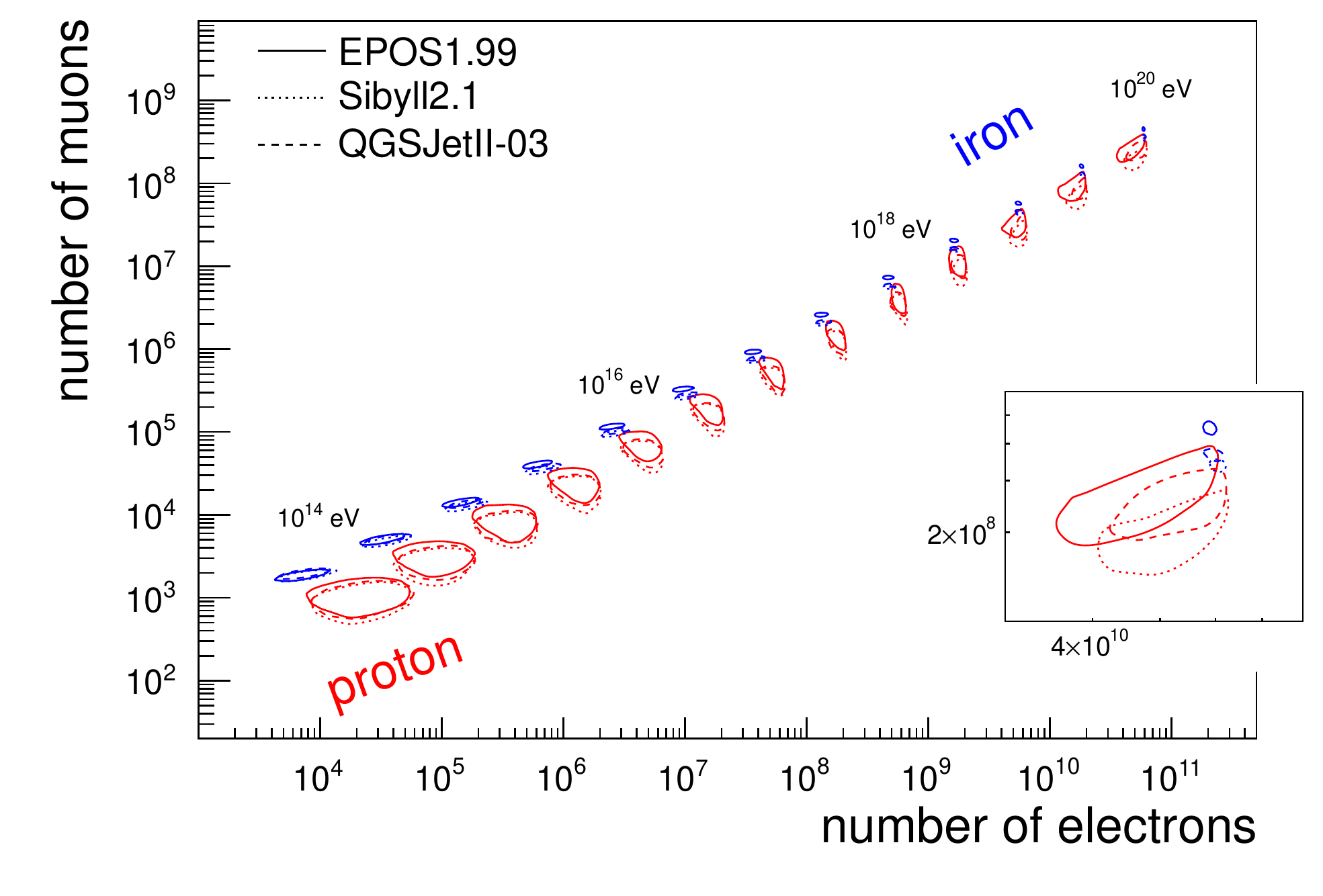}
  \caption[compoVariablesGround]{Air shower simulation
    of the number of muons vs.\
    electrons at ground level for a vertical shower observed at 
    800\,\gcm.
    Contour lines illustrate the regions which include 90\,\% of the 
    showers and
    the inset shows a detailed view at \energy{20}.}
  \label{fig:compoVariablesGround}
\end{figure*}
which demonstrates that $\sigma(\Xmax)^2$ is proportional to the
variance of $\ln A$.  Starting from this equation, Linsley proposed to
analyze air shower data in the $\sigma(\ln A)-\meanLnA$
plane~\cite{LinsleyXmax1985} which, however, involves the need to
specify the dependence of $\sigma_A$ on $\ln A$.  In this article we
will use a similar approach below (cf.\ Sec.~\ref{sec:fluoDets}), but
stick to the experimental $\sigma(\Xmax)$ and $\meanXmax$ values.  For
the purpose of comparing data at several energies to model
predictions, it is useful to subtract the (energy dependent) values
for iron nuclei from both the data and predictions. The approximate
energy independence of this subtracted variables is illustrated in
Fig.~\ref{fig:sigmaVsXmax}. Two-component transitions as predicted
from air shower simulations with {\scshape Sibyll} are shown as curves
that are, as expected from Eq.~(\ref{eq:sigmaTwoComp}), parabolas in
the $\sigma(\Xmax)$-$\meanXmax$ plane.  These parabolas define a
closed contour that contain {\itshape all possible} combinations of
mass mixtures for $A\in[1,56]$. To illustrate the discriminative power
of the $\sigma(\Xmax)$-$\meanXmax$ combination, three models for
energy evolution of the extragalactic cosmic ray
composition~\cite{Berezinsky:2002nc, Allard:2005cx, Allard-11} are
shown as well.

\subsection{Particles at Ground}
Another way of detecting cosmic rays and to estimate their mass is
given by the measurement of particle densities of air showers at
ground. In the calorimeter analogy of the previous section, this would
correspond to a calorimeter with only one active readout plane and
correspondingly this measurement technique is more susceptible to
shower-to-shower fluctuations. Nevertheless, ground measurements are
still frequently used in cosmic ray detectors because of their
geometric acceptance and high duty cycle.

An estimate of the qualitative dependencies of the number of muons and
electrons on primary mass and characteristics of hadronic interactions
can again be obtained within the Heitler model. Given the average
multiplicity $N$ of each interaction, the energy of charged and
neutral pions in a shower initiated by a primary proton of energy $E$
is $E_\pi = E/N^n$ after the $n$th interaction if one (somewhat
unrealistically) assumes an energy independent multiplicity. This
energy splitting continues until the charged pion energy reaches the
decay energy at which the hadronic interaction length
$\lambda_\mathrm{int}$ becomes equal to the decay length
$\lambda_\mathrm{dec} = \rho\,\gamma \,c \tau$, where $\rho$ is the
height-dependent density of air, $\gamma$ denotes the Lorentz-boost
and $\tau$ is the pion lifetime. For a shower with incident angle
$\theta$ in an isothermal atmosphere with scale height $h_0$, the
density at slant depth $X$ is
\begin{equation}
\rho(h)=\frac{X}{h_0}\cos\theta=\frac{n\lambda_\mathrm{int}}{h_0}\cos\theta.
\end{equation}
Therefore,
the condition $\lambda_\mathrm{int} = \lambda_\mathrm{dec}$ leads to a decay energy
that is independent of the interaction length. It is reached
after $n_\mathrm{d}$ interactions for which
\begin{equation}
n_\mathrm{d}\,N^{-n_\mathrm{d}} = \frac{h_0}{c\tau} \frac{m_\pi\,c^2}{E} \frac{1}{\cos\theta}
\label{eq:ndecay}
\end{equation}
and therefore
\begin{equation}
n_\mathrm{d} = - \frac{\mathrm{W}_{\kern -0.1em -1}\left(-\frac{h_0}{c\tau} \frac{m_\pi\,c^2}{E} \frac{\ln N}{\cos\theta}\right)}{\ln N},
\end{equation}
where $\mathrm{W}_{\kern -0.1em -1}$ denotes the lower branch of the Lambert-W function (see e.g.~\cite{darkoLambert}).
The decay energy is then given by
\begin{equation}
\EcritPi = \frac{E}{N^{n_\mathrm{d}}}
\end{equation}
for which we find numerical values of a few tens of~\GeV and a slow
decrease with primary energy in agreement with the estimates
of~\cite{Matthews:2005sd}.  The total number of muons produced in a
shower is equal to the number of pions with $E_\pi=\EcritPi$ and
therefore
\begin{equation}
  N_\mu^p \approx \left(\frac{E}{\EcritPi}\right)^\beta
\end{equation}
with
\begin{equation}
\beta = \frac{\ln \frac{2}{3} N} {\ln N},
\label{eq:beta}
\end{equation}
where the factor $\frac{2}{3}$ gives the approximate fraction of
charged pion secondaries.  Air shower simulations predict $\beta$ to
be in the range of 0.88 to 0.92~\cite{AlvarezMuniz:2002ne},
corresponding to effective multiplicities from 30 to 200 in
Eq.~(\ref{eq:beta}). It is interesting to note, that because the
interaction length drops out in the calculation of $n_{d}$ (cf.\
Eq.~(\ref{eq:ndecay})), the number of muons at ground are expected to
be independent of $\lambda_\mathrm{int}$.

The number of electrons at shower maximum, i.e.\ at the point at which
the electron energies become too low to produce new particles ($E_e =
\EcritEM$), can be estimated from the total amount of energy in the
electromagnetic cascade given by the primary energy minus the energy
in muons. Since $E_\mu = N_\mu\EcritPi$, the number of electrons is
\begin{equation}
N_{e, \mathrm{max}}^p = \frac{E}{\EcritEM}-\frac{\EcritPi}{\EcritEM}\left(\frac{E}{\EcritPi}\right)^\beta \approx \frac{E}{\EcritEM},
\end{equation}
where the last approximation can be made at high energies at which the
energy fraction transfered to muons becomes small.

Using again the superposition model and substituting $E$ with $E^\prime = E/A$,
one obtains the following relations for nuclear primaries:
\begin{equation}
N_{e, \mathrm{max}}^A \approx A\frac{E/A}{\EcritEM} = N_{e, \mathrm{max}}^p
\end{equation}
and
\begin{equation}
  N_\mu^A \approx A \left(\frac{E/A}{\EcritPi}\right)^\beta = N_{\mu, 
\mathrm{max}}^p\, A^{1-\beta}.
\end{equation}
So, whereas the number of electrons at shower maximum gives a good
estimate of the primary energy independent of the composition, the
number of muons can be used to infer the mass of the primary particle,
since it grows with $A^{1-\beta}$. Moreover, the evolution of the muon
number with energy, $\mathrm{d} N_\mu / \mathrm{d} \ln E$, is a good
tracer of changes in the primary composition.  Just as in the case of
the elongation rate of the longitudinal development, a constant
composition gives $\mathrm{d} N_\mu / \mathrm{d} \ln E = \beta$ and
any departure from that behavior can be interpreted as a change of the
average mass of the primaries.

Unfortunately, the experimental situation is more complicated, because
surface detectors do {\itshape not} observe the number of electrons at
shower maximum, but at a fixed depth
$X_\mathrm{ground}/\cos\theta$. If the detector and shower maximum are
separated by $\Delta X = X_\mathrm{ground}/\cos\theta - \Xmax$, then
only the attenuated number of electrons is observed with
\begin{equation}
  N_{e,\mathrm{ground}} \approx N_{e,\mathrm{max}}\, 
\exp{\left(-\frac{\Delta X}{\Lambda}\right)},
\label{eq:attenuation}
\end{equation}
where $\Lambda \approx 60$\,\gcm is the attenuation length of the
number of electrons after the shower maximum.  Since heavy primaries
reach their shower maximum at smaller depths than light ones, the
number of electrons on ground is expected to be composition sensitive
as well, with a larger electron number for air showers initiated by
light primaries.  This feature is visible in
Fig.~\ref{fig:compoVariablesGround}, where $N_\mu$ vs.\ $N_e$
is shown for air shower simulations at different energies for a
detector located at 800\,\gcm.  As can be seen, the $\ln N_\mu$ and $\ln
N_e$ observables are basically rotated from the desired
quantities, $\ln A$ and $\ln E$.  Due to the steeply falling cosmic
ray spectrum, this rotation causes a complication in the analysis of
air shower data, because showers of equal $\ln N_e$ are
enriched in light elements (cf.\ Sec.~\ref{sec:particleDet} for a
description of unfolding methods to overcome this
problem). Furthermore, Eq.~(\ref{eq:attenuation}) implies that given
the $\Xmax$ fluctuations explained in the last section, the relative
fluctuations of the electron number are expected to be quite
substantial,
\begin{equation}
  \frac{\sigma(N_{e,\mathrm{ground}})} {N_{e,\mathrm{ground}}} \approx
    \frac{\sigma(\Xmax)}{\Lambda}.
\end{equation}

These attenuation effects can be reduced considerably by choosing an
appropriate detector site which is situated at a height close to the
shower maximum.  The exponential attenuation
Eq.~(\ref{eq:attenuation}) is only valid far from the maximum, whereas
in its close vicinity the shower size is nearly invariant under small
displacements from the maximum (see Fig.~\ref{fig:longitudinal}
below). Since the simulations in Fig.~\ref{fig:compoVariablesGround}
were performed at a fixed ground depth of 800~\gcm, the evolution of
the attenuation effect with distance to the shower maximum can be seen
indirectly: At low energies where the observation level is far from
the shower maximum, the difference in the number of electrons between
proton and iron primaries is large and diminishes while the shower
maximum approaches the ground level at higher energies.

Besides the measurement of the number of electrons and muons,
experiments with surface detectors have further means to determine the
{\itshape shower age} (i.e.\ the distance to the shower maximum) by
studying the shape of the particle densities with respect to the
distance to the shower core. These measurements of the lateral
distribution as well as other additional composition sensitive
variables from ground detectors will be discussed
in~Sec.~\ref{sec:ldf}.

\begin{figure*}[ht!]
  \centering
  \includegraphics[width=0.7\linewidth]{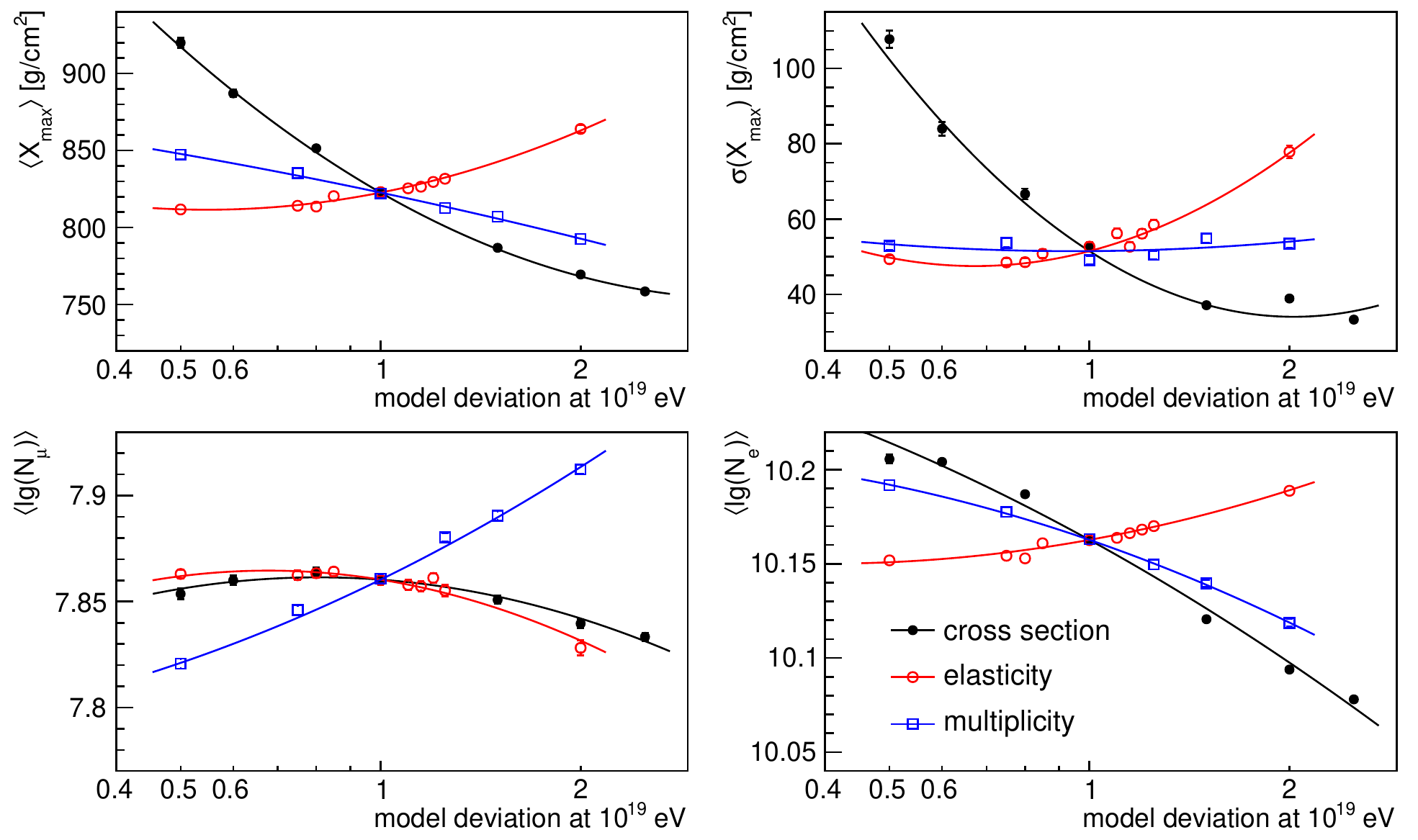}
  \caption[modelSys]{Sensitivity of air shower observables on
    characteristics of hadronic interactions as a function
    of a change in model characteristics growing logarithmically from zero at Tevatron energies to
    the quoted values on the x-axis at \energy{19} (proton showers at
    \energy{19.5}, adapted from~\cite{Ulrich:2010rg}). }
  \label{fig:modelSys}
\end{figure*}
\subsection{Model Uncertainties}
The physics of air showers is very well understood in terms of
particle transport through the atmosphere and for electromagnetic
showers it is currently believed that they can be modeled without any
significant uncertainties.  In the case of hadronic showers, however,
there is a fundamental lack of theoretical and experimental knowledge
of the characteristics of hadronic interactions (see e.g.\
\cite{EngelPierogHeck2011, Alessandro:2011wt} for recent discussions
of hadronic interactions in air showers).  Since most of the
interactions in an air shower are 'soft interactions', i.e.\ occur
with only a small momentum transfer, perturbative QCD is not
applicable.  The phenomenological interaction models that have to be
used instead, are constrained by low energy experimental data only,
but even at these energies the full phase space relevant for air
shower interactions is not fully covered~\cite{Heck:2003br,
  Drescher:2003gh,Meurer:2005dt, Maris:2009uc}.  At ultra-high
energies, the center of mass energies of the first nucleus-air
interactions are beyond accelerator energies and correspondingly the
models solely rely on extrapolations.

This somewhat bleak situation is currently alleviated by
special-purpose programs to measure particle production data relevant
for cosmic rays~(e.g.\ \cite{na61, Adriani:2008zz}) and by the wealth
of new data from the multipurpose detectors at the LHC with which the
interaction models can be tested and re-tuned.  At its maximum center
of mass energy of 14\,\TeV the LHC will eventually reach the equivalent
of \energy{17} in the laboratory system and thus cover most of the
energy range at which galactic cosmic rays are detected with air
showers.  First comparisons of the 7\,\TeV data to interaction models
used in air shower simulations suggest that the LHC data are bracketed
by these models~\cite{dEnterria:2011kw}.

The effect of uncertainties in hadronic interaction characteristics on
air shower observables has been recently studied
in~\cite{Ulrich:2010rg} and~\cite{Parsons:2011ad}.  Some of the
results from~\cite{Ulrich:2010rg} are displayed in
Fig.~\ref{fig:modelSys} where response of air shower observables to
ad-hoc changes of the {\scshape Sibyll} interaction model are shown.
As can be seen, all air shower observables are highly susceptible to
such changes but to a different degree. For instance, the muon number
and $\Xmax$ are very sensitive to the multiplicity (as expected from
Eqs.~(\ref{eq:protonXmax2}) and~(\ref{eq:beta}), whereas the $\Xmax$
fluctuations are not. And likewise there is a strong dependence of
$\Xmax$ and $\sigma(\Xmax)$ on $\lambda$ but not so for $N_\mu$ (as
suggested by $\lambda$-independence of $n_\mathrm{d}$ in
Eq.~(\ref{eq:ndecay})).

%% file: particleDet.tex
\begin{figure*}[ht!]
\centering
\subfigure[{\scshape QGSJet01}]{
\includegraphics[width=0.35\linewidth]{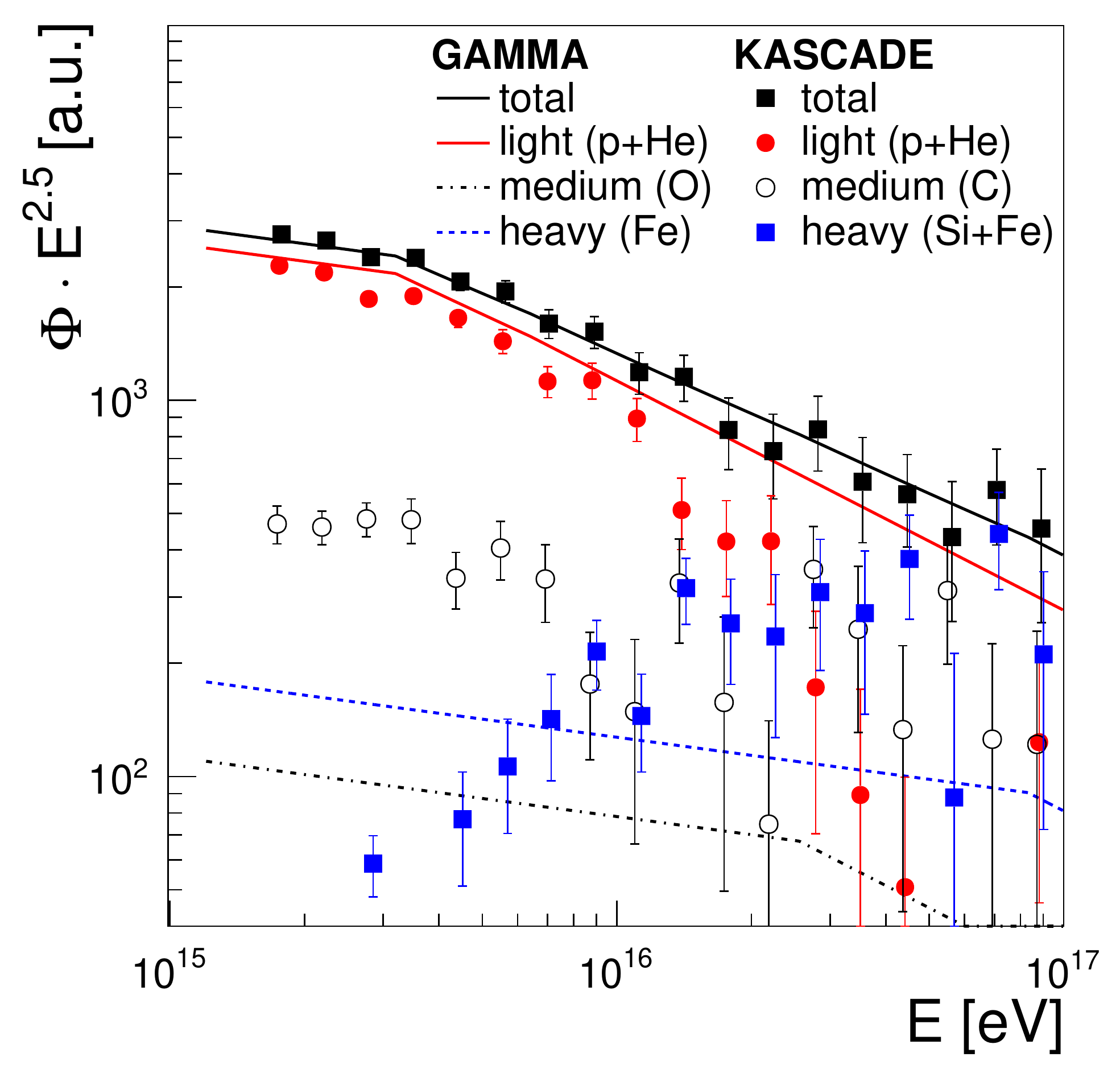}
\label{fig:FluxQ01}
}
\subfigure[{\scshape Sibyll2.1}]{
\includegraphics[width=0.35\linewidth]{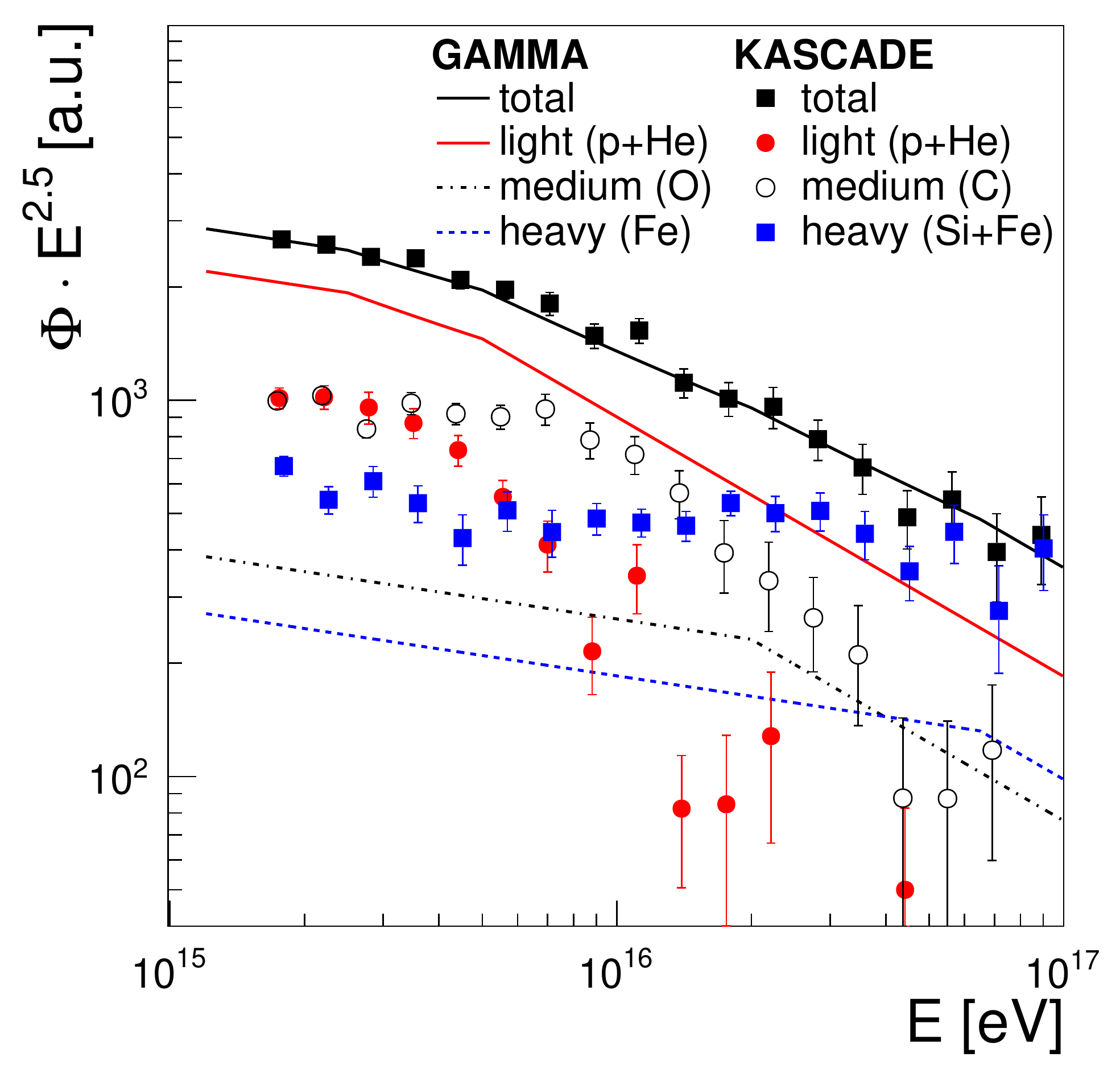}
\label{fig:FluxSib}
}
\caption[fluxes]{Unfolded fluxes from GAMMA~\cite{Garyaka-07} and
  KASCADE~\cite{KASCADE-05} using two different interaction
  models. The spectra of elemental groups have been adjusted by a common
  factor (within 35\%) to match the all-particle spectra of the two experiments. KASCADE
  points were slightly displaced in energy for better visibility.}
\label{fig:FluxPlot}
\end{figure*}

\subsubsection{$N_e$-$N_\mu$ Method}

Measuring electron and muon numbers (and their fluctuations) has
become the first and most commonly employed technique applied to infer
the cosmic ray composition from EAS data. The basic principles of this
approach can be understood very intuitively, since the sum of electron
and muon numbers measured at ground relates to the primary energy,
while the ratio of the muon to electron number relates to the primary
mass for reasons described in the previous section.  These general
features can also be observed in Fig.\,\ref{fig:compoVariablesGround}.
Fukui et al.~\cite{Fukui-60} and Khristiansen et
al.~\cite{Khristiansen-63} were the first to study the muon number
fluctuations in the knee region and they were also the first
concluding an enrichment by heavy nuclei above the knee
energy~\cite{Khristiansen94}. Similar conclusions were drawn
in~\cite{Elbert:1976tn} based on a larger data set.

In the most classical approach, electron-muon discrimination is
achieved by employing a combination of unshielded and shielded
scintillation detectors at ground level. Recent examples include
AGASA~\cite{Chiba:1992vr}, CASA-MIA~\cite{borione94},
EAS-TOP~\cite{Aglietta-93}, GRAPES~\cite{gupta-05},
KASCADE~\cite{kascade-03c}, KASCADE-Grande~\cite{Apel-NIM-10},
Maket-ANI~\cite{Danilova:1992wc}, GAMMA~\cite{Garyaka-07}, and
Yakutsk~\cite{yakutsk03}. The Pierre Auger Observatory~\cite{auger-04}
operates water Cherenkov detectors of 1.2~m depth which also enables
limited muon identification. The electromagnetic particles (electrons,
positrons, and photons) are more numerous than the muons but their
mean energy at detector level is only some 10\,\MeV while that of the
muons is about 1\,\GeV. The former thus produce a large number of
relatively small Cherenkov pulses whereas muons produce a small number
of large pulses. This type of discrimination technique was pioneered
at the Haverah Park detector~\cite{HP73}. Since FADC readout was not
available at that time, signal traces of the detectors were
photographically recorded from oscilloscopes. IceTop at the South
Pole~\cite{Achterberg-06b} uses tanks of frozen ice instead of water
for obvious reasons. Sensitivity to the primary composition is
achieved by measuring the dominant electromagnetic component at ground
level in IceTop in coincidence with high energy muon bundles in
IceCube (muon threshold about 500\,\GeV), originating from the first
interactions in the atmosphere. Similar combinations of surface and
underground detectors have been used e.g.\ by EAS-TOP and MACRO at the
Gran Sasso~\cite{eastop-04c} and at the Baksan underground
laboratory~\cite{bakatanov99}. Such methods are potentially very
attractive for composition studies, as the \TeV energy muons measured
deep underground probe the energy per nucleon while the surface array
probes the energy per particle of the cosmic ray primary. The
disadvantage is the reduced statistical accuracy of the muon
measurements and the limited solid angle available for coincident
cosmic ray observations.  Finally,
Tibet-AS$\gamma$~\cite{Amenomori-08A} complemented their air shower
array with emulsion chambers and so-called burst detectors, and
Telescope Array (TA)~\cite{TA-exp} employ solely unshielded
scintillation detectors and thus do not discriminate muons from
electrons at detector level. Composition studies in these cases rely
on additional measurements discussed below.

The importance and proper treatment of air shower fluctuations in any
analysis of the nuclear mass composition has already been
emphasized. It is important to realize that electron and muon numbers
do not fluctuate independently on event-by-event basis, but are
mutually correlated (cf.\ Fig.\,\ref{fig:compoVariablesGround}). All
these effects can be properly accounted for by a two-dimensional
unfolding technique, first utilized by the
KASCADE-Collaboration~\cite{KASCADE-05, Apel-09}. This approach yields
a set of energy spectra of primary mass groups, such that their
resulting simulated double differential electron and muon number
distribution, $\mathrm{d}^2N/\mathrm{d}N_e \mathrm{d}N_\mu$, resembles
the observed one. The only input required for solving the underlying
mathematical equations are the so-called kernel functions describing
the transformation $(E,A) \to (N_\mu^{\rm obs}, N_e^{\rm obs})$
including their shower-by-shower fluctuations. These functions need to
be generated from air shower simulations (including detector response
functions) and thus will depend on the chosen hadronic interaction
model employed in the EAS simulations. The level of systematic
uncertainties imposed by the interaction models can be studied by
constructing kernel functions using different hadronic interaction
models and comparing the corresponding unfolded results (cf.\ Fig.\
\ref{fig:FluxPlot}). Moreover, the observed two-dimensional
$(N_\mu^{\rm obs}, N_e^{\rm obs})$ distribution can be directly
compared to the forward folded distribution when using the unfolding
results as input.  Based on such kind of systematic studies the
KASCADE Collaboration concluded that neither \mbox{\scshape QGSJet-01}
nor {\scshape Sibyll2.1} is able to describe the full range of
energies consistently. Specifically, a muon deficit (or electron
excess) has been found for {\scshape Sibyll2.1} simulations when
compared to data~\cite{KASCADE-05}. Expressed in terms of the mean
logarithmic mass (see below) the two models differ systematically by
about half a unit but yield the same basic result of a light (He
dominated) composition at the knee with a change towards a heavy
composition at higher energies. Moreover, the data are consistent with
the assumption of a rigidity dependent change of the knee
energy. Similar conclusions about an increasing mass across the knee
have been drawn e.g.\
in~\cite{glasmacher99a,swordy02,eastop-04c}. More recently, the GAMMA
Collaboration~\cite{Garyaka-07} performed a forward folding approach
to their electron and muon distributions. In this case, specific
spectral shapes are to be defined (here: power-law indices below and
above the knee, $E_\text{k}/Z$, and the fluxes of four different mass
groups). By optimizing the spectral parameters one tries to match the
forward folded mean $N_e$ vs.\ $N_\mu$ distribution to the measured
data. The results of this analysis are included in
Fig.~\ref{fig:FluxPlot} and are shown as full and dashed lines. A good
description of the all-particle spectrum is again found for a light
composition at the knee with an increasing fraction of heavy primaries
above. However, differences between forward folding by {\scshape
  QGSJet-01} and {\scshape Sibyll2.1} lead to $\Delta \langle \ln A
\rangle \approx 0.7$-$0.8$ with an overall rather light composition
(cf.\ Fig.~\ref{fig:lnAsurface}). Data from Tibet
AS$\gamma$~\cite{Amenomori-06}, on the other hand, seem to suggest an
increasing mass composition already below the knee leading to a heavy
dominated composition at the knee. This conclusion, however, is based
on a number of assumptions and a rather limited statistics of
events. Since the set-up and event selection used in
Ref.~\cite{Amenomori-06} was essentially blind to heavy primaries, the
heavy component has been deduced in
Ref.~\cite{Amenomori-06,Amenomori:2011hu} rather indirectly by
subtracting the flux of the light component (p+He) from the
all-particle spectrum measured by Tibet-III~\cite{Amenomori-03}.

\begin{figure}[t]
  \includegraphics[width=\linewidth]{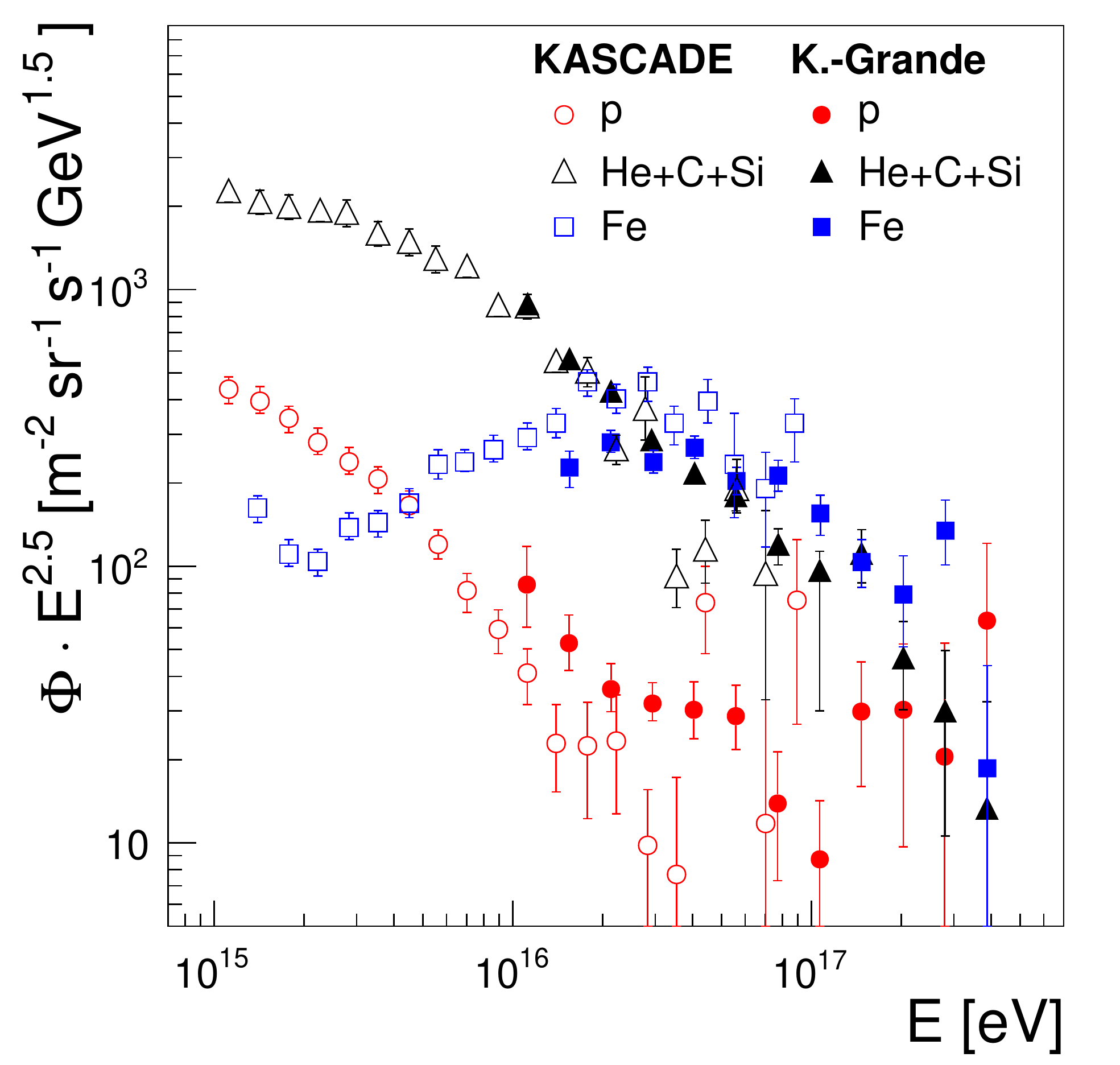}
  \caption[fluxKASC]{Unfolded fluxes from KASCADE and KASCADE-Grande
    (open and full symbols, respectively) based on the {\scshape QGSJet-II-02}
    interaction model (data are from~\cite{HaungsICRC11,
      FuhrmannICRC11}.\label{fig:fluxKA-Grande}}
\end{figure}

Only very recently, data from the KASCADE-Grande experiment reached
sufficient statistics allowing the unfolding analyses to be extended
to energies nearing \energy{18} (cf.\ Fig.~\ref{fig:fluxKA-Grande}).
Because of poorer detector resolution in KASCADE-Grande, the analysis
has been limited to only three mass groups. However, the preliminary
unfolding results \cite{FuhrmannICRC11} confirm earlier findings of
KASCADE and indicate a very heavy composition at about
\energy{17}. Similar conclusions about the mass composition using
electron-poor and electron-rich showers have been drawn
in~\cite{Apel:2011mi} with a knee-like structure observed at
$8\times10^{16}$\,\eV that could be identified with the heavy component
of primary cosmic rays.

\subsubsection{Steepness of Lateral Distribution Function}
\label{sec:ldf}
The determination of the lateral density distribution of particles in
EAS is the most fundamental task of a particle detector, because the
total number of charged particles, electrons, or muons is obtained by
fitting the radial fall-off of the measured particle densities to
proper analytical functions and performing the radial integration. The
energy and mass of the primary particle can then be deduced from the
electron and muon numbers, as described in the previous
section. Historically, choices of parameterizations of both electron
and muon lateral distributions were influenced very much by the
seminal review of Greisen~\cite{greisen60} in which the analytical
calculations performed by Kamata and Nishimura~\cite{Kamata-60} for
electromagnetic showers were generalized to the
Nishimura-Kamata-Greisen (NKG) function:
\begin{eqnarray}
\rho_{\rm NKG}(r,s,N_e) & = &
  \frac{N_e}{r_M^2} \;\;
  \frac{\Gamma(4.5-s)}{2\pi\Gamma(s)\Gamma(4.5-2s)} \nonumber \\
  & & \times \; \Bigg(\frac{r}{r_M}\Bigg)^{s-2}
  \Bigg(1+\frac{r}{r_M}\Bigg)^{s-4.5}
\label{eq:NKG}
\end{eqnarray}
with the shower age $0.5 < s < 1.5$, the Moli{\`e}re radius $r_M$, and
the electron shower size $N_e$. A large variety of Lateral
Distribution Functions (LDFs) is found in the literature (named
Greisen, Greisen-Linsley, etc.), often being specific modifications of
the NKG-function aiming at providing an optimum description for either
electrons or muons, or both, or are simple power-laws being optimized
to describing densities measured by scintillators or water Cherenkov
detectors, etc.~\cite{Coy:1997wu,kascade-01a,Lagutin:2002vq}.

Besides being a tool to reconstruct shower sizes, the actual shape of
the LDF also contains information about the underlying particle
physics in the EAS and, thereby, also about the mass of the primary
particle. Generally, showers initiated by heavy primaries and reaching
their maximum at high altitudes will exhibit a flatter LDF than those
initiated by light primaries and developing deeper into the
atmosphere. This feature is observed both for electrons and muons and
the steepness of the LDF can be extracted for samples of EAS (properly
selected by shower size and zenith angle)~\cite{ave-02a} or, if
sampling statistics does allow, also on event by event
basis~\cite{kascade-01a}. A reanalysis of data from Volcano
Ranch~\cite{ldf-linsley} was performed in Ref.~\cite{Dova-04} and
yielded an iron fraction of about 75\% at $E\simeq 10^{18}$\,eV, if
the {\scshape QGSJet-01} model was adopted. The sensitivity of the LDF
to the primary mass, however, is weaker as compared to the
$N_e$-$N_\mu$ method or those techniques discussed below.

\subsubsection{Muon Tracking and Timing}
Interest in reconstructing the the mean heights of production of muons
in EAS to learn about the longitudinal shower development date back to
the 60ies of the last century~\cite{Earnshaw:1973un}. However, at that
time neither the angular resolution of EAS arrays nor the accuracy of
shower simulations were appropriate to the use such measurements for
investigating the mass composition of primary cosmic rays. The
technique was revived in the 90ies with the Cosmic Ray Tracking (CRT)
detectors at HEGRA (aiming at tracking of electrons {\em and\/}
muons)~\cite{bernloehr96}, the Muon Tracking Detector at
KASCADE~\cite{doll-02} or at GRAPES~\cite{Hayashi-05}. In these
detectors the orientation of the muon track is measured with respect
to the shower axis and the muon production height is then
reconstructed by means of triangulation. For such detectors one needs
to find a compromise between a sufficiently long base-line at the
ground (i.e.\ distance between shower core and tracking detector) and
a sufficiently large area for the tracking detectors. Increasing the
average distance improves the accuracy of the triangulation but shifts
the tracking detector in regions of smaller muon densities.  A typical
compromise at energies around and above the knee is about 100\,m
baseline or more and at least 100\,m$^2$ muon tracking area. Based on
such measurements, no or only a small rise of the mean mass above the
knee was concluded from CRT at HEGRA~\cite{bernloehr98}. However, data
were binned as a function of the electron shower size which imposes a
strong bias towards electron rich showers, i.e.\ to light
primaries. Measurements of the muon production height with
KASCADE-Grande~\cite{Apel:2011cr} are compatible with a clear
transition from light to heavy cosmic ray primary particles with
increasing shower energy across the knee.

An attractive alternative to infer the mean muon production depth in
EAS was suggested in~\cite{Cazon:2004in,Cazon:2005gl}. Instead of
performing a geometrical triangulation by complex and expensive
tracking detectors, one measures the time delay of muons in each
detector station at ground level with respect to the shower
front. Since muons are in general produced close to the shower axis,
those muons produced at high altitude can reach a detector station at
some distance to the shower core at a shorter average path length as
compared to muons produced deep in the atmosphere. Therefore, the
relative time delay to the shower front plane is a direct measure of
the mean Muon Production Depth (MPD) in the atmosphere. Similarly to
the aforementioned muon tracking detectors, the resolution of
measurements of the MPD improves both with increasing muon numbers
observed in a detector station and with increasing distance of the
station to the shower core. However, distortions become large close to
the shower core because of limited FADC sampling times and electron
contamination. Thus, a proper minimum distance to the shower core
needs to be maintained.

A first application of this novel method to real data has been
presented by the Pierre Auger Collaboration only very
recently~\cite{GarciaGamez:2011}. From the MPD distribution measured
in each event, the quantity $X_\mu^{max}$ can be extracted by fitting
a Gaisser-Hillas function~\cite{ghfunc} to this distribution, where
$X_\mu^{max}$ can be interpreted as the point where the production of
muons reaches the maximum along the cascade development.  The
resolution in individual events ranges from about 120\,\gcm at the
lower energies to less than 50\,\gcm at the highest energy and the
systematic uncertainties on $\langle X_\mu^{max}\rangle$ were reported
11\,\gcm. Within these uncertainties, a mixed composition with a mild
change to a heavier composition is reported in~\cite{GarciaGamez:2011}
for energies above $2\cdot 10^{19}$ (cf.\ Fig.~\ref{fig:lnAsurface}).

\subsubsection{Rise-Time}

Further information about the primary mass of UHECR is encoded in the
time profile of shower particles. This was first noted by Bassi et
al.~\cite{Bassi:1953vl} and was used in the context of primary
composition at Haverah
Park~\cite{Watson:1974wq,Walker:1981vv,Walker:1982vp}. An advantage of
this technique is that it does not require electron-muon
discrimination at the detector level, but makes use of the fact that
muons are mostly detected near the shower front and that electrons are
subject to stronger attenuation in the atmosphere than muons. Full
exploitation of this technique, however, requires state-of-the-art
FADC read-out electronics. Moreover, it has been noted that the
signals from water Cherenkov detectors located upstream of an inclined
shower exhibit faster rise-times than those located downstream. This
is called the `rise-time asymmetry'~\cite{Dova:2009ew}. As rise-time
we shall understand here the time it takes for the integral signal to
increase from 10 to 50\% of its full value.

For simple geometrical reasons, no rise-time asymmetry is expected for
vertical showers. For very inclined showers one again finds only very
small rise-time asymmetries between inner and outer detectors. This
can be understood because of the absence of the electromagnetic
component and dominance of the high energy muon component in inclined
showers, the latter being basically free of asymmetries. The zenith
angle at which the rise-time asymmetry becomes maximum,
$\theta_{max}$, was demonstrated to be correlated with the shower
development~\cite{Dova:2009ew} and thereby to the average mass of the
primary particles. Like the $N_e$-$N_\mu$ unfolding, this method
cannot be applied on an event-by-event basis, which however is no
major limitation. A first application to data was presented in
Ref.~\cite{GarciaPinto:2011} indicating a transition from a light to
heavy composition in the energy range $3\times10^{18}$ to $4\times
10^{19}$\,eV (cf.\ Fig.~\ref{fig:lnAsurface}).

%% file: nicDets.tex
\begin{figure}[t!]
 \includegraphics[width=\linewidth]{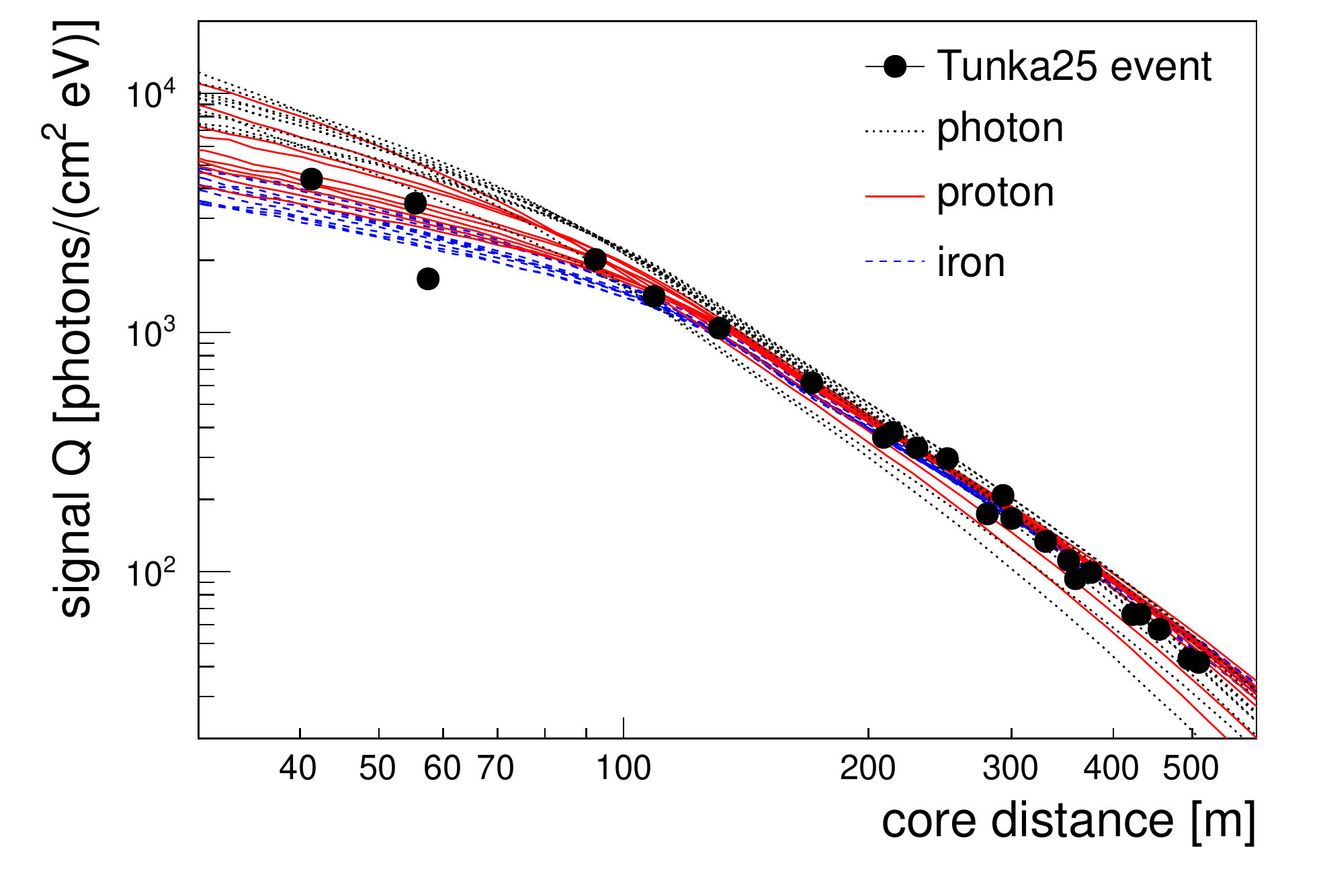}
 \caption[exampleLongi]{Example of a lateral Cherenkov light
    distribution as measured with non-imaging Cherenkov
    detectors. Data points are taken from~\cite{Budnev:2009mc}
    ($E=0.12$\,\EeV) and compared to ten simulated~\cite{soft:CORSIKA}
    air showers for three different primary particle types using the
    hadronic interaction model {\scshape
    QGSJetII}~\cite{Ostapchenko:2010vb}.\label{fig:Tunka-LDF}}
\end{figure}

If atmospheric conditions permit, particle measurements at ground are
often complemented by observations of Cherenkov or fluorescence light
from EAS.

Cherenkov light emitted by extensive air showers in the atmosphere has
been known for many years to contain information on shower development
\cite{Brennan:1958wi}. Non-imaging observations by operating PMTs with
large Winston cones looking upwards into the night-sky are perhaps the
simplest and most straightforward technique of EAS observations. The
basic principles have been worked out in~\cite{Patterson:1983uq,
  hillasLateralCher1982} and the method was first successfully applied
at energies around the knee by the AIROBICC detectors installed at the
HEGRA array \cite{karle95} and by CASA-BLANCA~\cite{fowler2001} in
Utah. More recent measurements at the Tunka \cite{Budnev:2009mc,
  Antokhonov:2011fy} and Yakutsk \cite{Knurenko:2011} arrays have
extended these measurements up to ultra-high energies.

\begin{figure*}[t!]
  \includegraphics[height=.39\linewidth]{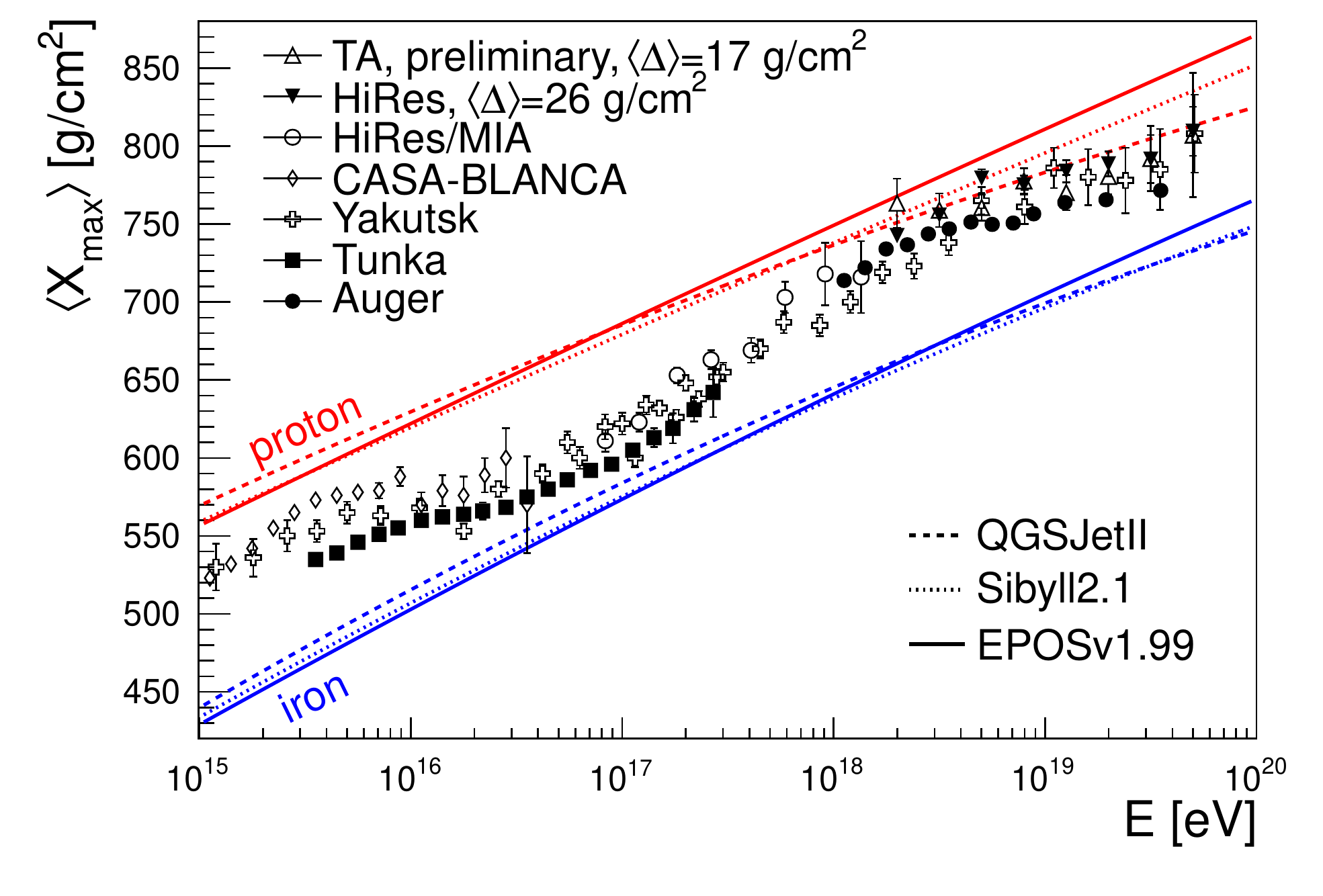}
  \includegraphics[height=.39\linewidth]{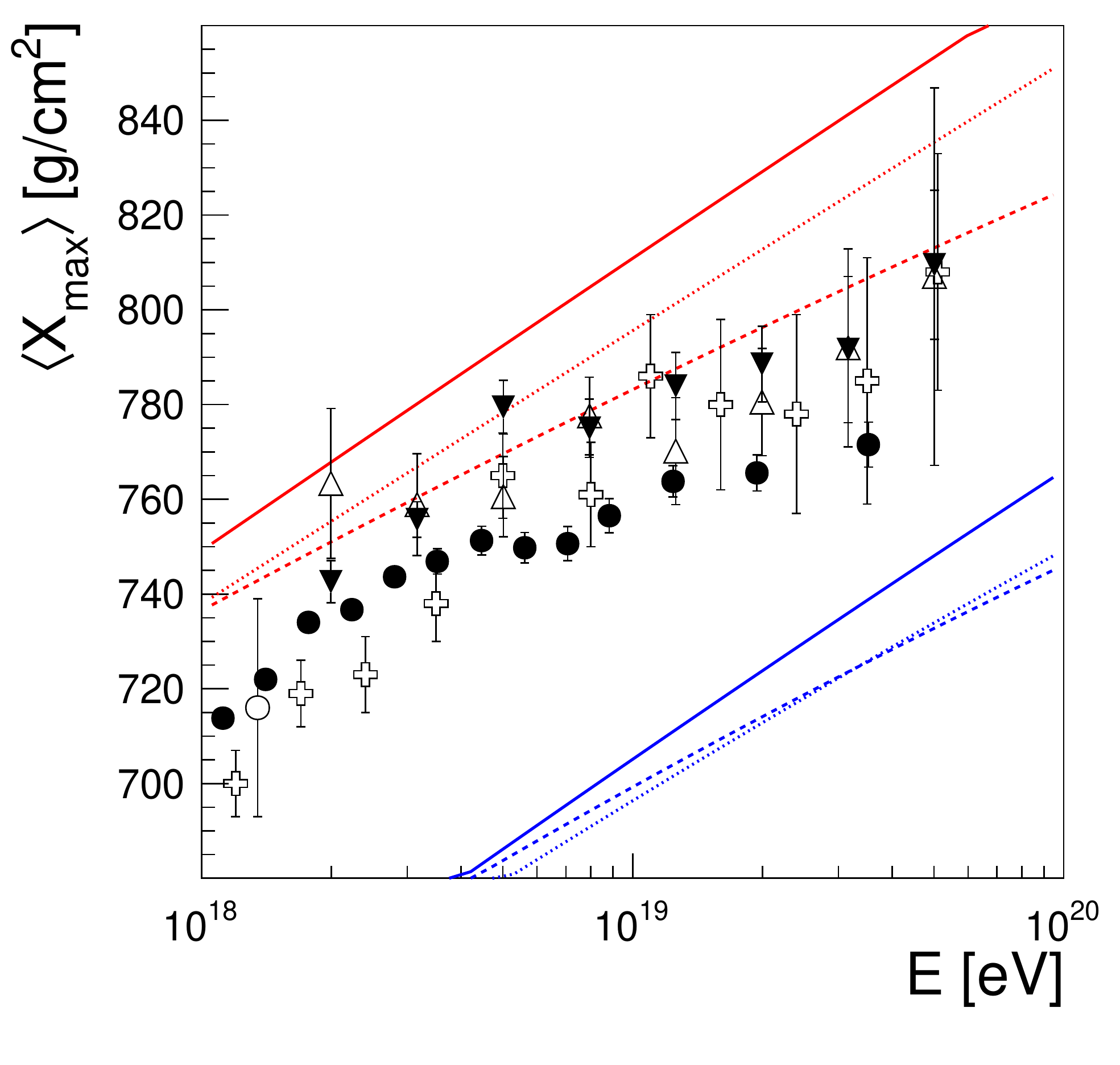}
  \caption[xmax]{Measurements of \meanXmax with non-imaging Cherenkov
    detectors (Tunka~\cite{Budnev:2009mc, ProsinICRC11},
    Yakutsk~\cite{Knurenko:2010eu, Knurenko:2011zz},
    CASA-BLANCA~\cite{fowler2001}) and fluorescence detectors
    (HiRes/MIA~\cite{AbuZayyad:2000ay}, HiRes~\cite{Abbasi:2009nf},
    Auger~\cite{AugerXmaxICRC11} and TA~\cite{Jui:2011vm}) compared to
    air shower simulations~\cite{Bergmann:2006yz} using hadronic
    interaction models~\cite{Pierog:2006qv, Ahn:2009wx,
      Ostapchenko:2010vb}. HiRes and TA data have been corrected for
    detector effects as indicated by the $\langle\Delta\rangle$ values
    (see text). The right panel shows a zoom to the ultra-high energy
    region.}
  \label{fig:xmax}
\end{figure*}

The lateral distribution of Cherenkov light at ground is the result of
a convolution of the longitudinal profile of charged particles in the
shower with the energy threshold for Cherenkov emission and the
Cherenkov emission angle which both depend on the air density and thus
height. Moreover, the angular distribution of electrons in a shower
due to multiple scattering contributes to the lateral extend of
Cherenkov light at ground.  Because the electron energy distribution
(and thus the number of electrons above Cherenkov threshold) as well
as their angular distribution are universal in shower
age~\cite{Hillas:1982vn, Giller:2004cf,Nerling:2005fj,
  Lafebre:2009en}, the non-imaging Cherenkov technique provides a
model independent method to measure both, the calorimetric energy and
shower maximum of air showers.

A characteristic feature of the lateral light distribution at ground is
a prominent shoulder at around 120~m from the shower core
(cf.\ Fig.~\ref{fig:Tunka-LDF}) which is due to the strongly forward
beamed emission of the Cherenkov light ($\theta_\text{Ch}^\text{\,air}
\approx 1.4^\circ$) from near the shower maximum in the
atmosphere. The slope of the lateral distribution measured within this
120~m is found to depend on the height of the shower maximum and hence
on the mass of the primary cosmic ray nucleus. The overall Cherenkov
intensity at distances beyond the shoulder, on the other hand, is
closely related to the calorimetric energy.

The \meanXmax measurements from BLANCA~\cite{fowler2001},
Tunka~\cite{Budnev:2009mc, ProsinICRC11} and Yakutsk
\cite{Knurenko:2011} are shown in Fig.~\ref{fig:xmax}.  At low
energies ($E<10^{16}$\,\eV) the three measurements disagree by up to
40\,\gcm, but all three detectors observed small elongation rates above
$5\times10^{15}$\,\eV, indicating a change towards a heavier
composition. At around \energy{17} the absolute values of \meanXmax
from Tunka and Yakutsk are approaching the simulation results for
heavy primaries and beyond that energy the average shower maximum
increases again towards the air shower predictions for light
primaries.  At even higher energies, only the Yakutsk array measured
$\meanXmax$ with Cherenkov detectors and we will discuss this range in
the next section together with the data from fluorescence telescopes.

%% file: fluoDets.tex
After the first prototyping and detection of fluorescence light from
air showers~\cite{Porter:1970et, hara1977, Bergeson:1977nw}, the Fly's
Eye detector~\cite{Baltrusaitis:1985mx} and its successor
HiRes~\cite{AbuZayyad:2000uu} established the measurement of the longitudinal
development of air showers using fluorescence telescopes and studied
the evolution of the shower maximum with energy~\cite{flysEyeXmax1990,
  Abbasi:2004nz}. Currently, two observatories are in operation
that use the fluorescence technique for the determination of the
energy scale and for composition studies: The Pierre Auger Observatory
in the Southern hemisphere~\cite{Abraham:2009pm} and the Telescope
Array (TA) in the Northern hemisphere~\cite{TA-exp}.

The measurement of the longitudinal air shower development with
fluorescence telescopes relies on the fact that the charged
secondaries of an air shower excite the nitrogen molecules in the
atmosphere that in turn emit fluorescence light.  Since the light
yields~\cite{fluoWorkshop08} are proportional to the energy deposited
in the atmosphere, this observation allows to reconstruct the
longitudinal development of the air shower as a function of slant
depth.

\begin{figure}[t!]
  \includegraphics[width=\linewidth]{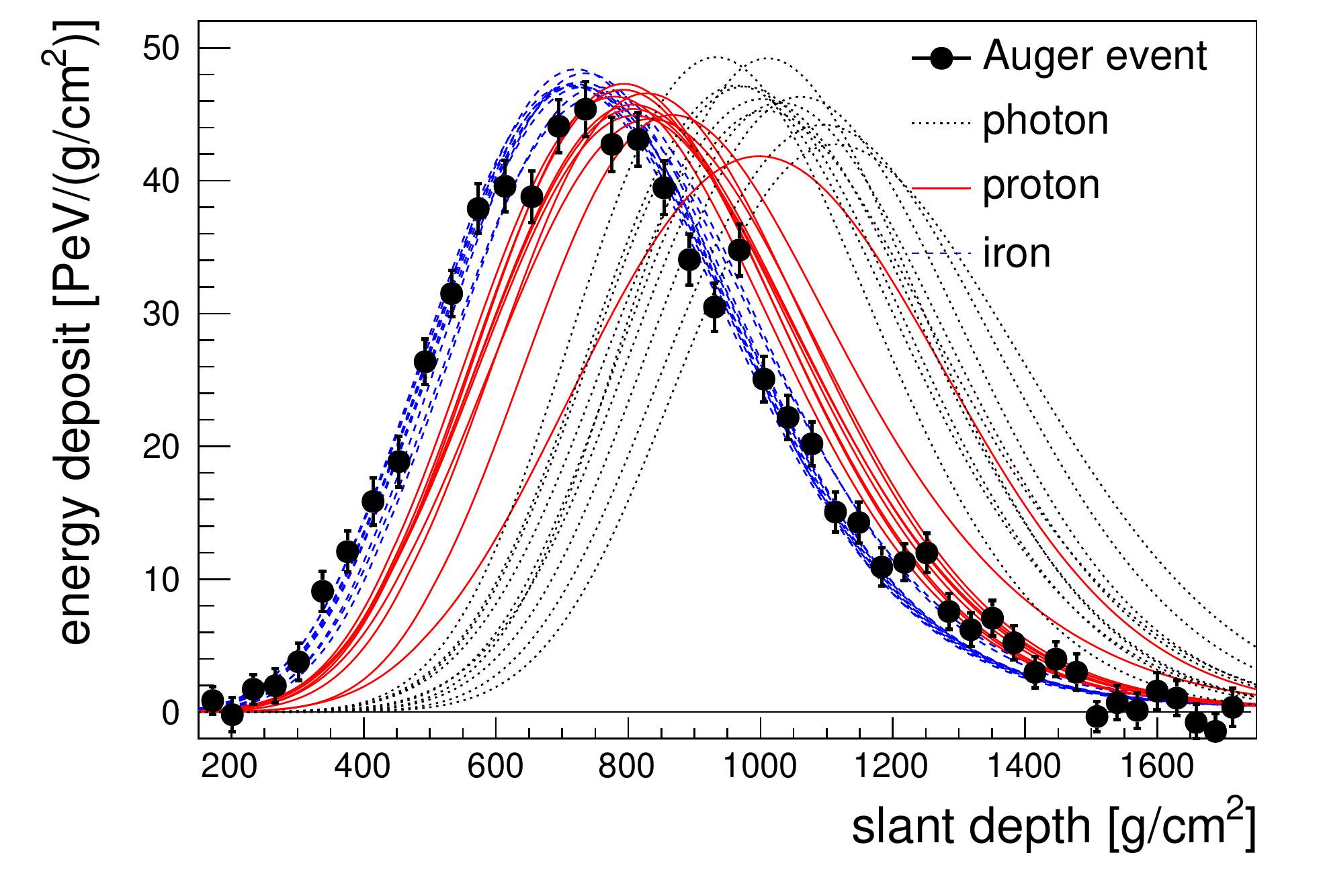}
  \caption[exampleLongi]{Example of a longitudinal air shower
    development as measured with fluorescence telescopes. Data points
    are taken from~\cite{Abraham:2009pm}
    ($E=(30\pm2)$\,\EeV) and compared to ten
    simulated~\cite{Bergmann:2006yz} air showers for three different
    primary particle types using the hadronic interaction model {\scshape
      Epos1.99}~\cite{Pierog:2006qv}.}
  \label{fig:longitudinal}
\end{figure}

A typical example of a reconstructed energy deposit profile of an
ultra-high energy air shower is shown in Fig.~\ref{fig:longitudinal}.
For this particular shower, the full profile was observed and the
total calorimetric energy could be obtained by simply adding up the
data points. In general, however, only part of the profile can
be detected, because the shower either reaches ground or its rising edge
is obscured by the upper field of view boundary of telescope.
Therefore, the profile is usually fitted with an appropriate trial
function~\cite{linsleyLongi} that allows the extrapolation of the
shower outside of the field of view and to below ground level. Popular
choices for fitting longitudinal profiles are the Gaisser-Hillas
function~\cite{ghfunc} (used e.g.\ by Auger~\cite{Unger:2008uq}) or a
Gaussian in shower age~\cite{AbuZayyad:2000np} as it was used for the
final HiRes analyses. The calorimetric energy of the shower is then
given by the integral of the fitted energy deposit profile.

In addition to the calorimetric energy, the measurement of the
longitudinal energy deposit profile provides a direct observation of
the shower maximum. As can be seen in Fig.~\ref{fig:longitudinal},
where simulated longitudinal shower profiles are superimposed on the
measured profile, even on a shower-by-shower basis a rough distinction
between heavy and light primaries is possible by comparing the
position of $\Xmax$.  In principle, the full distribution of shower
maxima for showers with similar energy contains the maximum
information about composition that can be obtained from fluorescence
detectors. Given enough statistics and an exact knowledge of the
expected distributions for different primaries, it should be possible
to extract composition groups (see e.g.~\cite{Abbasi:2006vp}) similar
to what is done for surface detectors. In the following, however, we
will concentrate on the first two moments of the $\Xmax$-distribution,
\meanXmax and \sigmaXmax.

For the determination of the average shower maximum, experiments bin
the recorded events in energy and calculate the mean of the measured
shower maxima.  For this averaging not all events are used, but only
those that fulfill certain quality requirements that vary
from experiment to experiment, but all analyses accept only profiles
for which the shower maximum had been observed within the field of
view of the experiment. Without this condition, one would rely only on
the rising or falling edge of the profile to determine its maximum,
which was found to be to unreliable to obtain the precise location of
the shower maximum. The field of view of fluorescence
telescopes is typically limited to 1-30 degrees in elevation. Therefore some
slant depths can only be detected with smaller efficiencies than others,
resulting in a distortion of the measured $\Xmax$-distribution due to
undersampling in the tails of the distribution~\cite{Unger:2009kk,
  Bellido:2009gm}. For instance, a detector located at a height
corresponding to 800\,\gcm vertical depth cannot detect shower maxima
deeper than 800, 924 and 1600\,\gcm for showers with zenith angles of
0, 30 and 60 degrees respectively. On top of this {\itshape acceptance
  bias} an additional {\itshape reconstruction bias} may be present
that can further distort the measured \meanXmax-values.

There are two ways to deal with such biases: If one is only interested
in comparing the data to air shower simulations for different primary
particles, then the biased data can be simply compared to air shower
predictions that include the experimental distortions. For this
purpose the full measurement process has to be simulated including the
attenuation in the atmosphere, detector response and reconstruction to obtain
a prediction of the observed average shower maximum, $\meanXmax_\mathrm{obs}$.
Another possibility is to restrict the data sample to shower
geometries for which the acceptance bias is small (e.g.\ by discarding
vertical showers) and to correct the remaining reconstruction effects to obtain
an unbiased measurement of \meanXmax in the atmosphere.

Whereas the former approach maximizes the data statistics, the latter allows
the direct comparison of published data to air shower simulations even for
models that were not developed at the time of publication. Moreover,
only measurements that are independent of the detector-specific distortions
due to acceptance and reconstruction can be compared directly.

\begin{figure}[t!]
  \includegraphics[width=\linewidth]{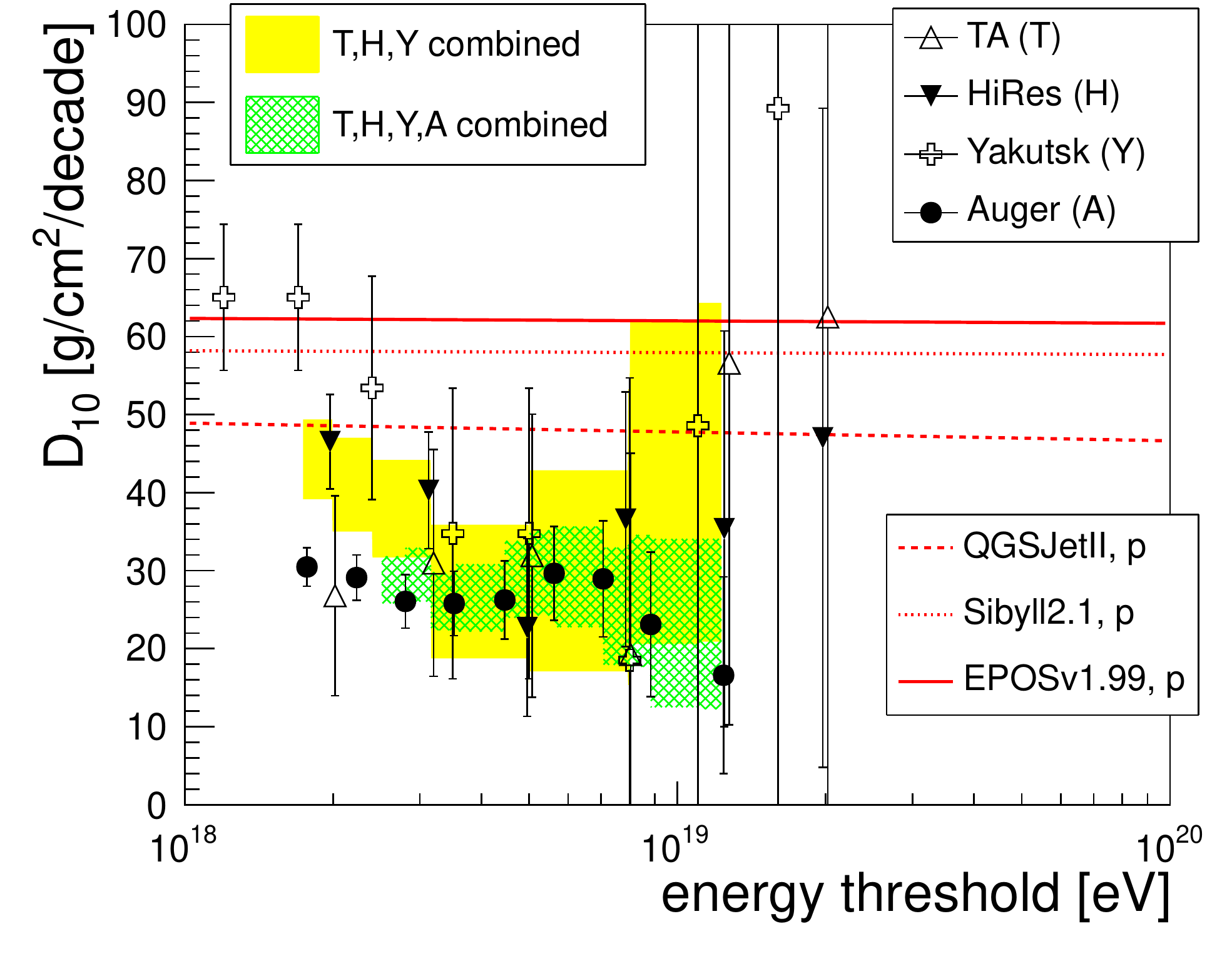}
  \caption[elongationRate]{Elongation rates obtained by a linear fit
    in $\lg E$ to the \Xmax data of HiRes, Yakutsk, TA and Auger above
    different energy thresholds. Only fit results with
    $\chi^2/\mathrm{Ndf}<2$ are shown. The yellow, solid band is the
    average obtained for HiRes, Yakutsk and TA , the green hatched
    band indicates the average for all four experiments.}
  \label{fig:elongationRate}
\end{figure}
The HiRes and TA collaborations follow the strategy to publish
  $\meanXmax_\mathrm{obs}$~\cite{Abbasi:2009nf, Jui:2011vm} and to
  compare it to the detector-folded air shower simulations. In the
  HiRes analysis the cuts were optimized to assure an \Xmax-bias that
  is constant with energy, but different for different primaries and
  hadronic interaction models. The preliminary TA analysis uses only
  minimal cuts resulting in energy dependent detection biases.  The
  Auger collaboration quotes average shower maxima that are without
  detector distortions within the quoted systematic
  uncertainties~\cite{Abraham:2010yv} due to the use of fiducial
  volume cuts. Yakutsk derives \Xmax indirectly using a relation
  between the slope of the Cherenkov-LDF and height of the shower
  maximum (cf.\ Sec.~\ref{sec:nicDets}). This relation is derived from
  air shower simulations and is universal with respect to the
  primary particle and hadronic interaction
  models~\cite{Korosteleva:2007ek}. We will therefore assume in the
  following, that the the Yakutsk measurement is bias-free and that it can be
  compared to air shower simulations directly.
\begin{figure*}[t!]
   \centering
  \includegraphics[width=0.8\linewidth]{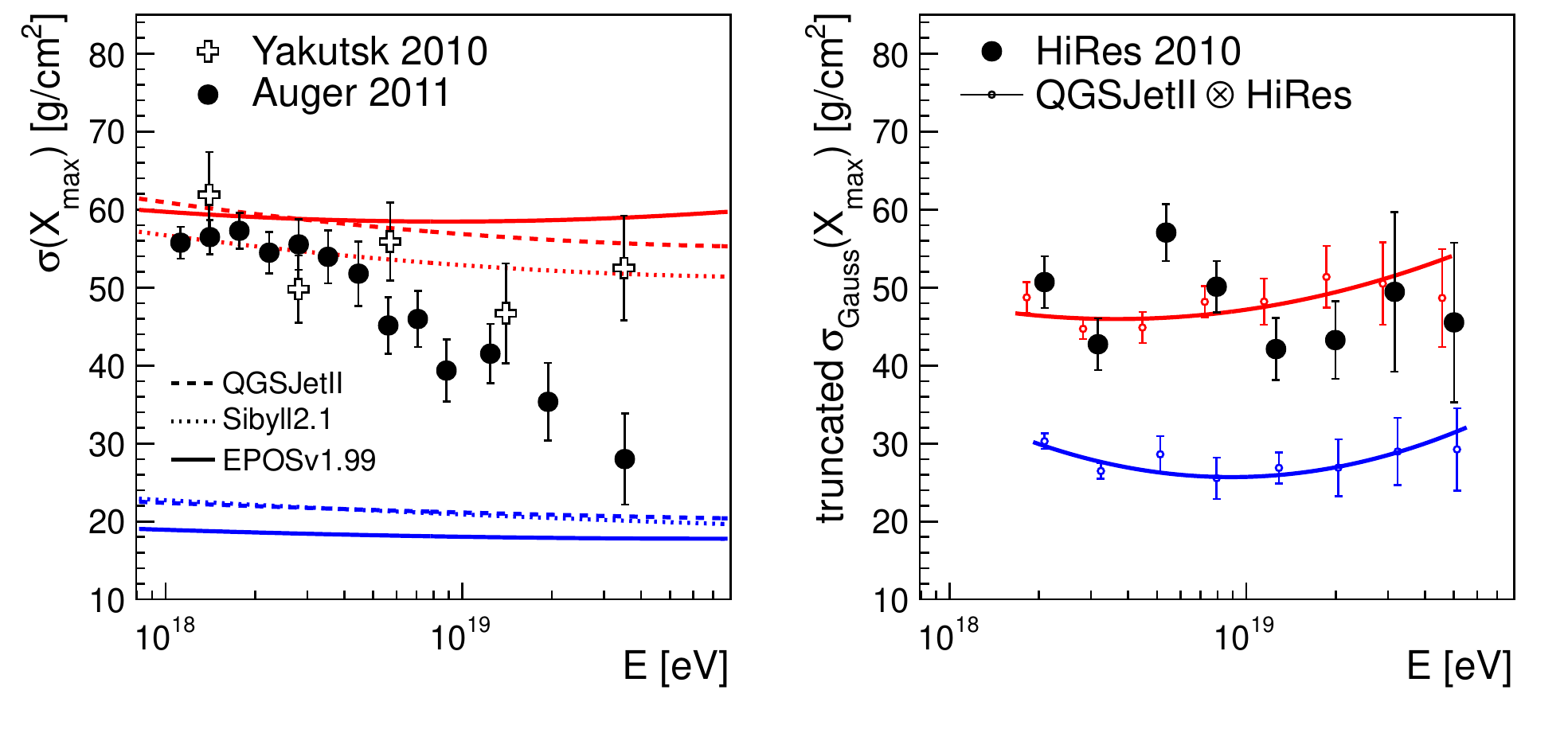}
  \caption[rmsPlot]{Left: Shower-to-shower fluctuations of $\Xmax$ in the atmosphere
   as measured by Yakutsk~\cite{Knurenko:2010eu} and Auger~\cite{AugerXmaxICRC11}
   compared to air shower simulations~\cite{Bergmann:2006yz}.
   Right: Width of a Gaussian fit to the truncated $\Xmax$-distributions
   as measured by HiRes~\cite{Abbasi:2009nf} compared to air shower simulations
    including detector effects.}
   \label{fig:rmsplot}
\end{figure*}

To allow a comparison of the
results of these experiments and moreover to calculate \meanLnA using
the {\scshape Epos} model (cf.\ Sec.~\ref{sec:lnA}) which was not
used in some of the original publications, we correct the
$\meanXmax_\mathrm{obs}$-values of HiRes and TA by shifting them by an
amount $\Delta$ which we infer from the difference of the published
$\meanXmax_\mathrm{obs}$-values for proton, {\scshape QGSJetII} to the
simulated values that are obtained without detector distortions:
\begin{equation}
  \meanXmax_\mathrm{corr} =  \langle X_\mathrm{max}\rangle_\mathrm{obs} + \Delta
\end{equation}
where
\begin{equation}
 \Delta = \left\langle X_\mathrm{max}(\textsc{QGSJetII}, \mathrm{p})\right\rangle -
\left\langle X_\mathrm{max}(\textsc{QGSJetII}, \mathrm{p})\right\rangle_\mathrm{obs}.
\end{equation}
The obtained $\Delta$-values are approximately constant with energy for
HiRes with $\langle\Delta\rangle=26$\,\gcm.
$\Delta$ decreases with energy for TA from $\Delta=24$\,\gcm at
\energy{18.3} to $\Delta=9$\,\gcm at \energy{19.7}.  Although these
values correspond to up to 25\% of the predicted proton/iron
difference, the correction procedure seems justified, because both
experiments measured a $\meanXmax_\mathrm{obs}$ that is very close to
the detector-distorted \textsc{QGSJetII}/proton predictions and the
$\Delta$-correction preserves the residuals wrt.\ this prediction for
$\meanXmax_\mathrm{corr}$.

The \meanXmax-values measurements from fluorescence detectors are
shown in Fig.~\ref{fig:xmax}. At around the region of the ankle of the
cosmic ray spectrum the measurements are compatible within their
quoted systematic uncertainties ($\sim$10\,\gcm) and the $\meanXmax$ is
close to the prediction for air showers initiated by a predominantly
light composition. Below this energy, the few data points from Auger
as well as the HiRes/MIA data confirm the trend observed by the
non-imaging Cherenkov detectors, namely a large elongation rate
indicative of a transition from a heavy to a lighter composition.  A
complete coverage of the transition region down to \energy{17} with
new fluorescence detector measurements is expected in the near future
with the High Elevation Auger Telescopes (HEAT, operating since
2009)~\cite{mathesICRC11} and the Telescope Array Low Energy Extension
(TALE, under construction)~\cite{thomsonICRC11}.

At ultra-high energies, the experimental situation can be quantified
by fitting the data with a linear function in the logarithm of energy.
The resulting elongation rates are
summarized in Fig.~\ref{fig:elongationRate}. As can be seen, the elongation
rates of the four experiments agree within uncertainties above
a few \EeV, though the statistical precision deteriorates quickly with
increasing energy.
Numerical values for fits above \energy{18.6}
are given Tab.~\ref{tab:XmaxFit}. Auger,
which has collected by far the largest statistics of the four
experiments, observes a small elongation rate of $D_{10} =26\pm
5$\,(\gcm)/decade, which could indicate a gradual increase of the
average mass of cosmic rays at ultra-high energies if the elongation
rate for a constant composition is indeed 50 to 60\,(\gcm)/decade as
predicted by the models.
%(31.9177/pow(18.1253,2)+22.7092/pow(11.9256,2)+34.7679/
% pow(18.6291,2))/(1./pow(18.1253,2)+1./pow(11.9256,2)+1./pow(18.6291,2)) = 2.755e+01
%sqrt(1./(1./pow(18.1253,2)+1./pow(11.9256,2)+1./pow(18.6291,2))) = 8.785
This small value of $D_{10}$ is confirmed by HiRes, TA, and Yakutsk,
for which the weighted average of elongation rates from Tab.~\ref{tab:XmaxFit}
is 27$\pm$9\,(\gcm)/decade.

\begin{table}[b!]
  \centering
  \begin{tabular}{lccc}
  \toprule
            & $X_{19}$  &   $D_{10}$           & {$\chi^2$/Ndf}\\
            & [$\unit{g/cm^2}$]    &  [(\gcm)/decade]    &             
  \\
  \midrule
    HiRes   & 782$\pm$3 & 23$\pm$11  &1.7/4\\
    Yakutsk & 773$\pm$5 & 35$\pm$19           &1.9/5\\
    Auger   & 758$\pm$1 & 26$\pm$5\phantom{0} &1.9/5\\
    TA      & 774$\pm$5 & 32$\pm$18           & 1.4/4\\
  \bottomrule
  \end{tabular}
  \caption{Results of a fit with 
    $\meanXmax=X_{19}+D_{10}(\lg(E/\text{eV}) -19)$
    to $\meanXmax$-data above \energy{18.6}.}
  \label{tab:XmaxFit}
\end{table}

\begin{figure*}[t!]
\centering
\subfigure[{\scshape QGSJet01}]{
\includegraphics[width=0.25\linewidth]{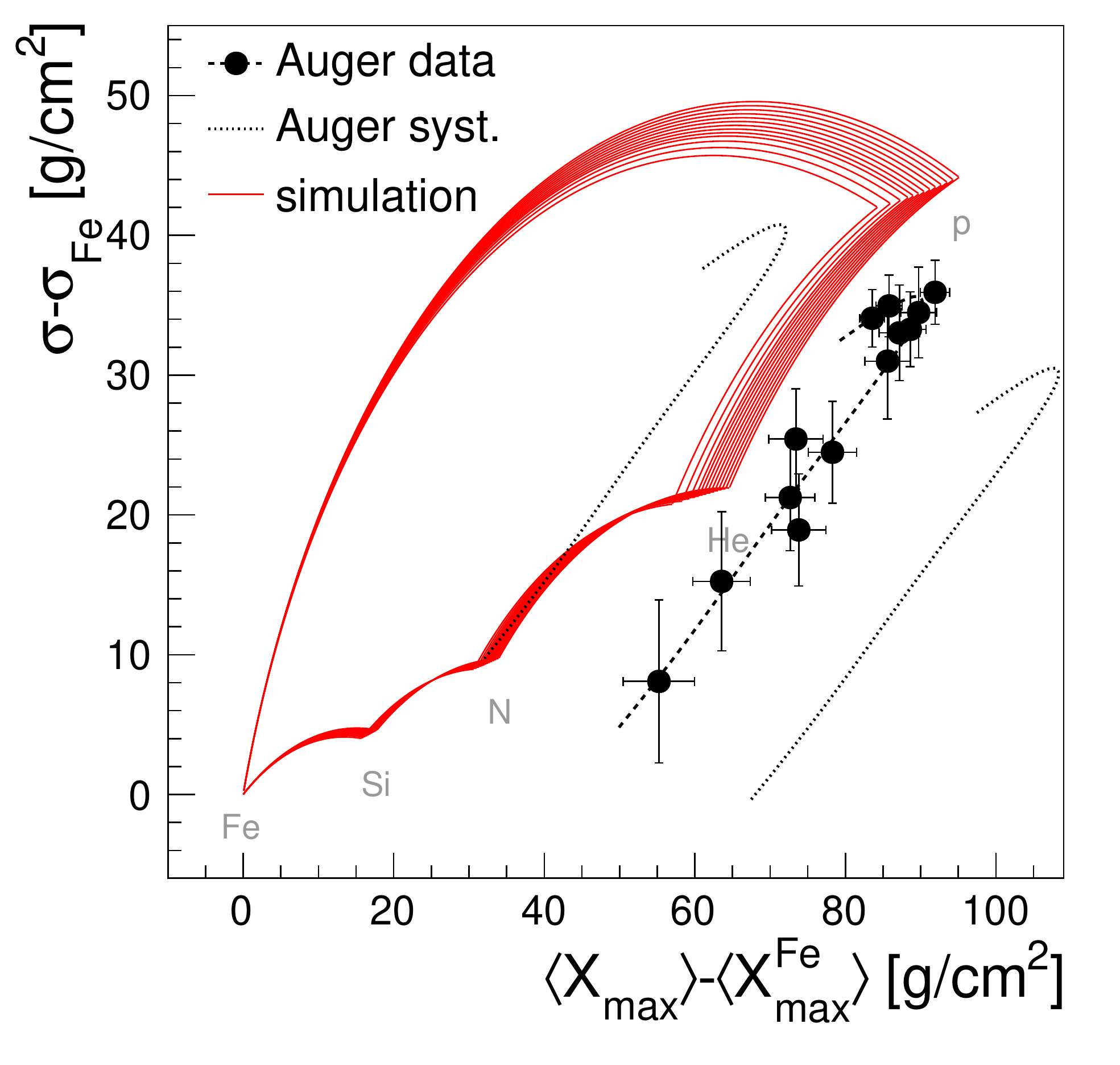}
\label{fig:augerUmbrella1}
}\hspace*{-0.2cm}
\subfigure[{\scshape QGSJetII}]{
\includegraphics[width=0.25\linewidth]{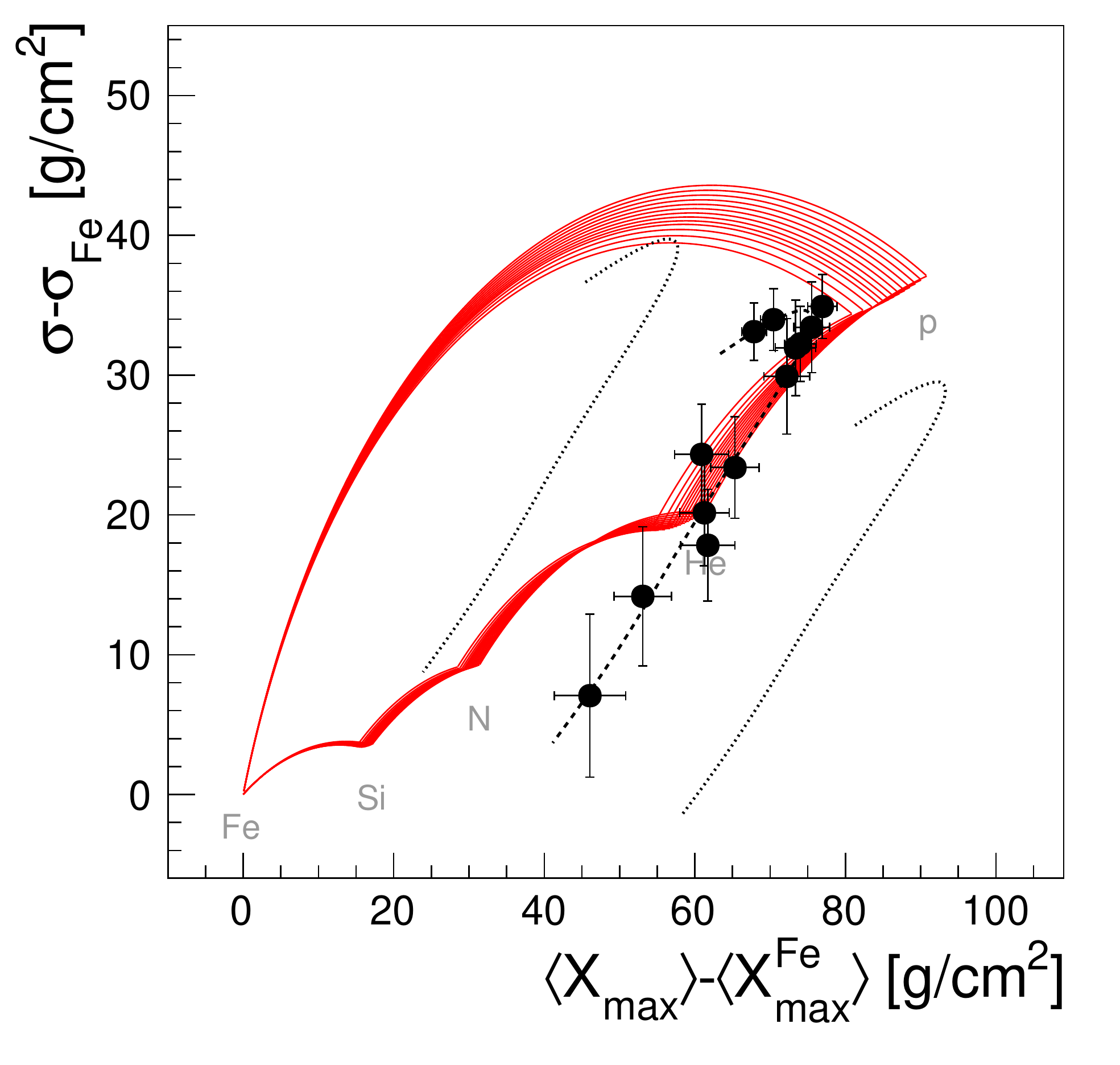}
\label{fig:augerUmbrella2}
}\hspace*{-0.2cm}
\subfigure[{\scshape Sibyll2.1}]{
\includegraphics[width=0.25\linewidth]{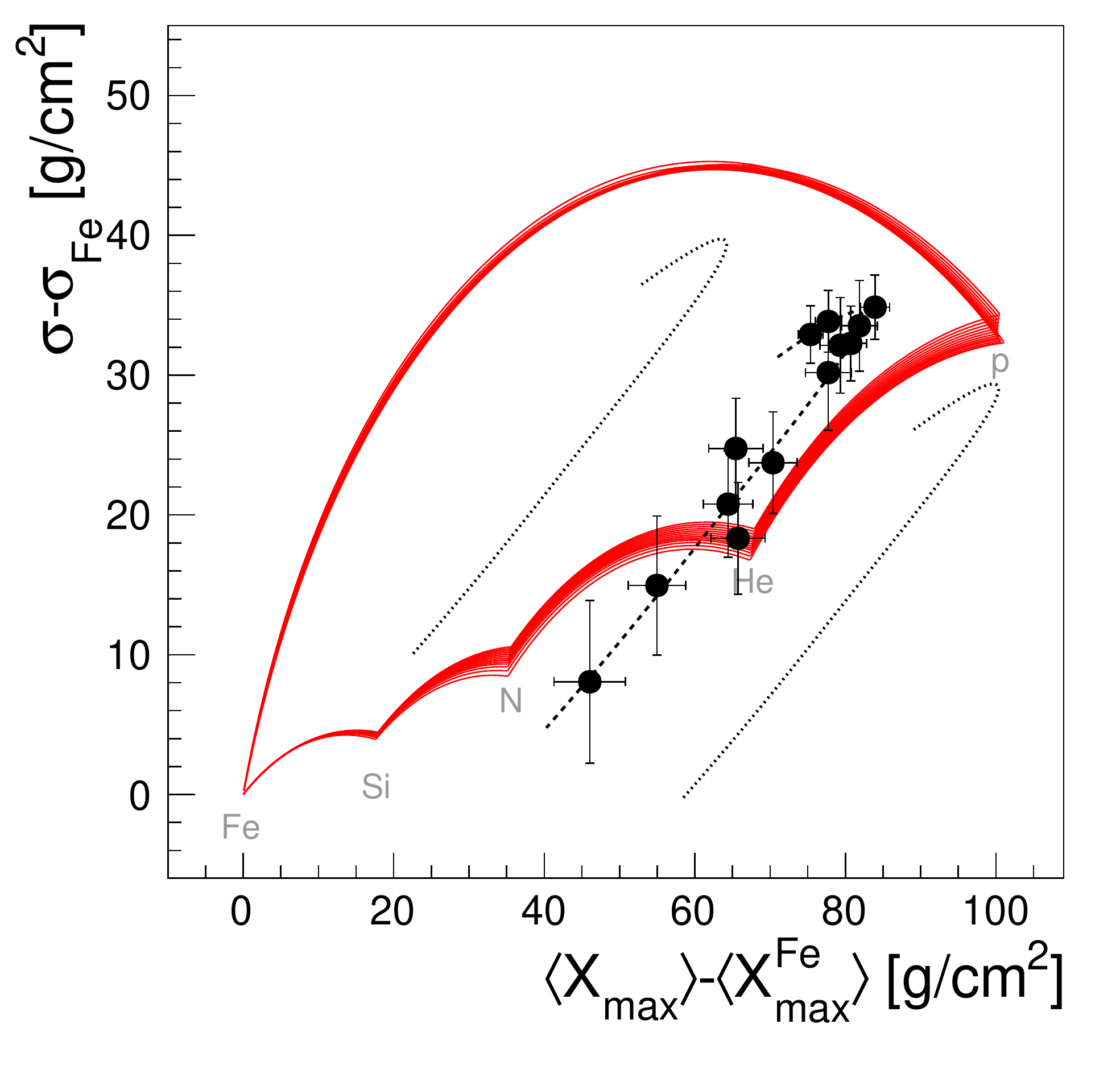}
\label{fig:augerUmbrella3}
}\hspace*{-0.2cm}
\subfigure[{\scshape Epos1.99}]{
\includegraphics[width=0.25\linewidth]{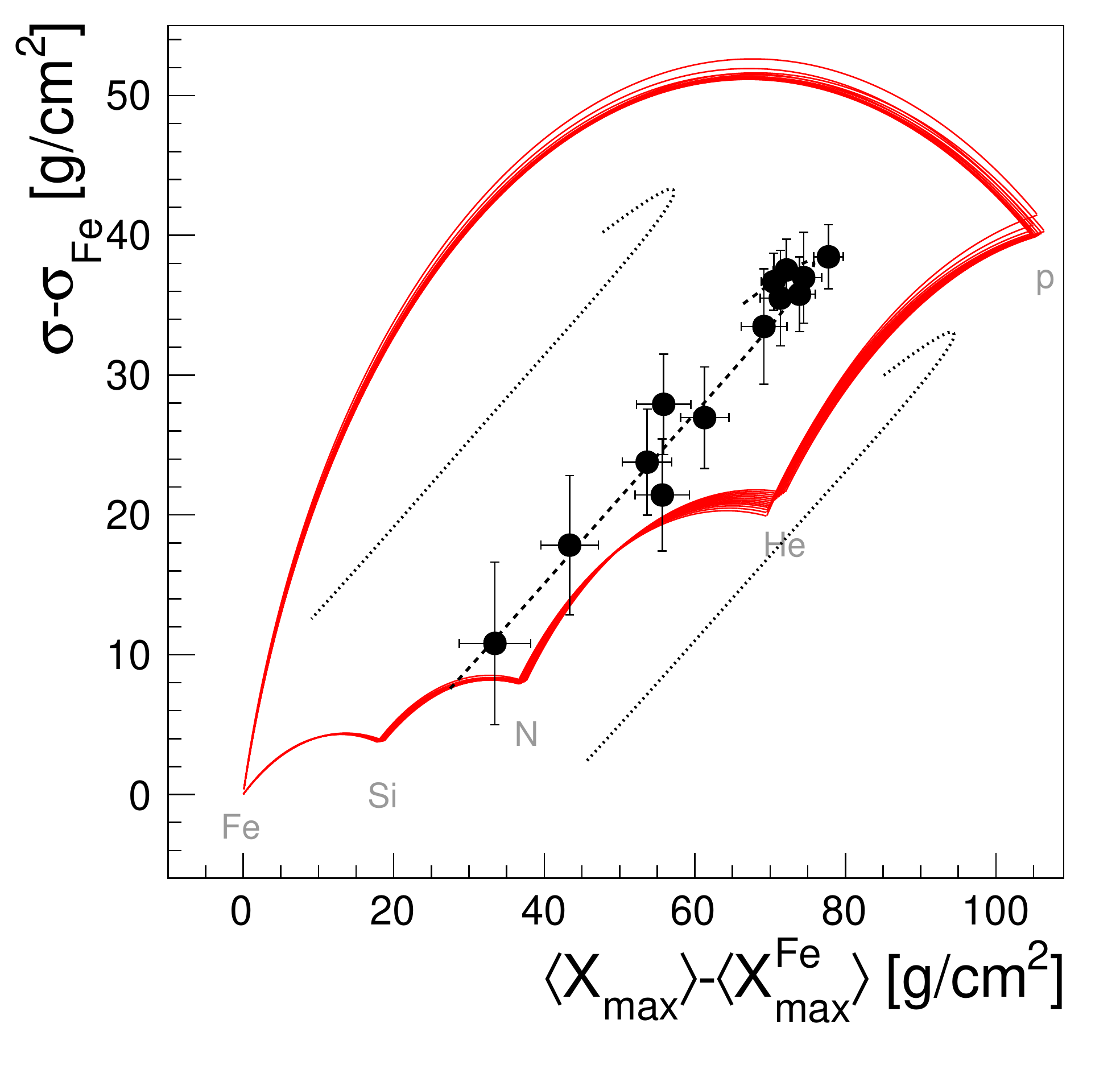}
\label{fig:augerUmbrella4}
}\hspace*{-0.2cm}
\caption[Auger umbrella]{Comparison of Auger measurements~\cite{AugerXmaxICRC11} of
                         the average shower depth and its fluctuations with air
                         shower simulations. The energy of the data points increases
                         from top to bottom, as indicated by the dashed line.\label{fig:augerUmbrella}}
\end{figure*}

The absolute depths at \energy{19} are, however, in poor agreement
among the four experiments and the differences of up to 24\,\gcm
between the Auger and HiRes results are larger than expected for 
individual
systematic uncertainties at the 10\,\gcm level. It is worthwhile noting
that {\itshape without} the $\Delta$ correction, the three fluorescence
detectors agree almost perfectly at ultra-high energies.  This might be a mere
coincidence or a hint that either HiRes and TA overestimate their bias
or that the assumption of $\Delta\approx0$ does not hold for Auger.

As explained in Sec.~\ref{sec:longDev}, the fluctuations of the shower
maximum provide another composition sensitive observable. The
measurements of \sigmaXmax from Auger and Yakutsk are shown in the
left panel of Fig.~\ref{fig:rmsplot}. Both data sets were corrected
for the detector resolution and can thus be directly compared to air
shower simulations. The Auger data exhibits a significant narrowing of
the \Xmax distributions with energy starting at about the energy of
the ankle. The low energy width is compatible with a light or mixed
composition, but the small \sigmaXmax at high energies points to the
presence of a significant fraction of CNO or heavier nuclei with
little admixture of light nuclei (cf.\ Eq.~(\ref{eq:sigmaMixed})). The
data of the Yakutsk array are compatible with a light composition at
all energies. Below \energy{19.3} there is a good agreement with the
Auger results, but their last data point is in contradiction with the small width
quoted by Auger.

The fluctuation measurements of HiRes are shown in the right panel of
Fig.~\ref{fig:rmsplot}. Instead of \sigmaXmax, HiRes published the
width of a Gaussian fit to the \Xmax-distributions that were truncated
at $\pm2\sigmaXmax$ without correction for detector resolution. This
variable can then be compared to air shower simulations including
detector effects.  As can be seen, HiRes finds a large width at low
energy that is, similar to Auger and Yakutsk, compatible with a light
or mixed composition. Above \energy{19} the width remains compatible
with proton simulations, albeit with large statistical uncertainties
that could also accommodate a narrower width.

The compatibility of the \meanXmax and \sigmaXmax measurements
with air shower simulations can be studied within the
$\sigma(\Xmax)$-$\meanXmax$ plane introduced in
Sec.~\ref{sec:longDev}. The Yakutsk data would cluster at around
the simulated proton values in this plane and the HiRes data cannot
be analyzed in this way without a full detector simulation.
Therefore only the Auger data are shown in Fig.~\ref{fig:augerUmbrella}.

If hadronic interactions at ultra-high energies are modeled correctly
and if the cosmic ray composition is any mixture of elements between
proton and iron, then the data points must lie within the contours
shown in Fig.~\ref{fig:augerUmbrella}.  As can be seen, this is not
the case for the outdated {\scshape QGSJet01} model. For both,
{\scshape QGSJetII} and {\scshape Sibyll2.1}, the Auger data are
barely at the edge of the contour, which would imply a somewhat
unnatural transition from a proton- to helium- to nitrogen-dominated
composition.  Using {\scshape Epos1.99}, the corresponding composition
would be mixed at low energies and very nitrogen-rich at high
energies. Whereas these considerations clearly demonstrate the power
of a combined observation of \meanXmax and \sigmaXmax, the current
systematic uncertainties of the Auger measurements do not allow for a
stringent test of the models: If the Auger data are shifted
simultaneously by $\pm\mathrm{syst}(E)$, $\mp\mathrm{sys}(\meanXmax)$
and $\pm\mathrm{syst}(\sigmaXmax)$ as in~\cite{ungerUHECRNagoya}
(indicated by lines in Fig.~\ref{fig:augerUmbrella}), then even for
{\scshape QGSJet01} a marginal compatibility remains.

%% file: lnA.tex
\begin{figure*}[ht!]
\centering
\subfigure[{\scshape QGSJet01}]{
\includegraphics[width=0.45\linewidth]{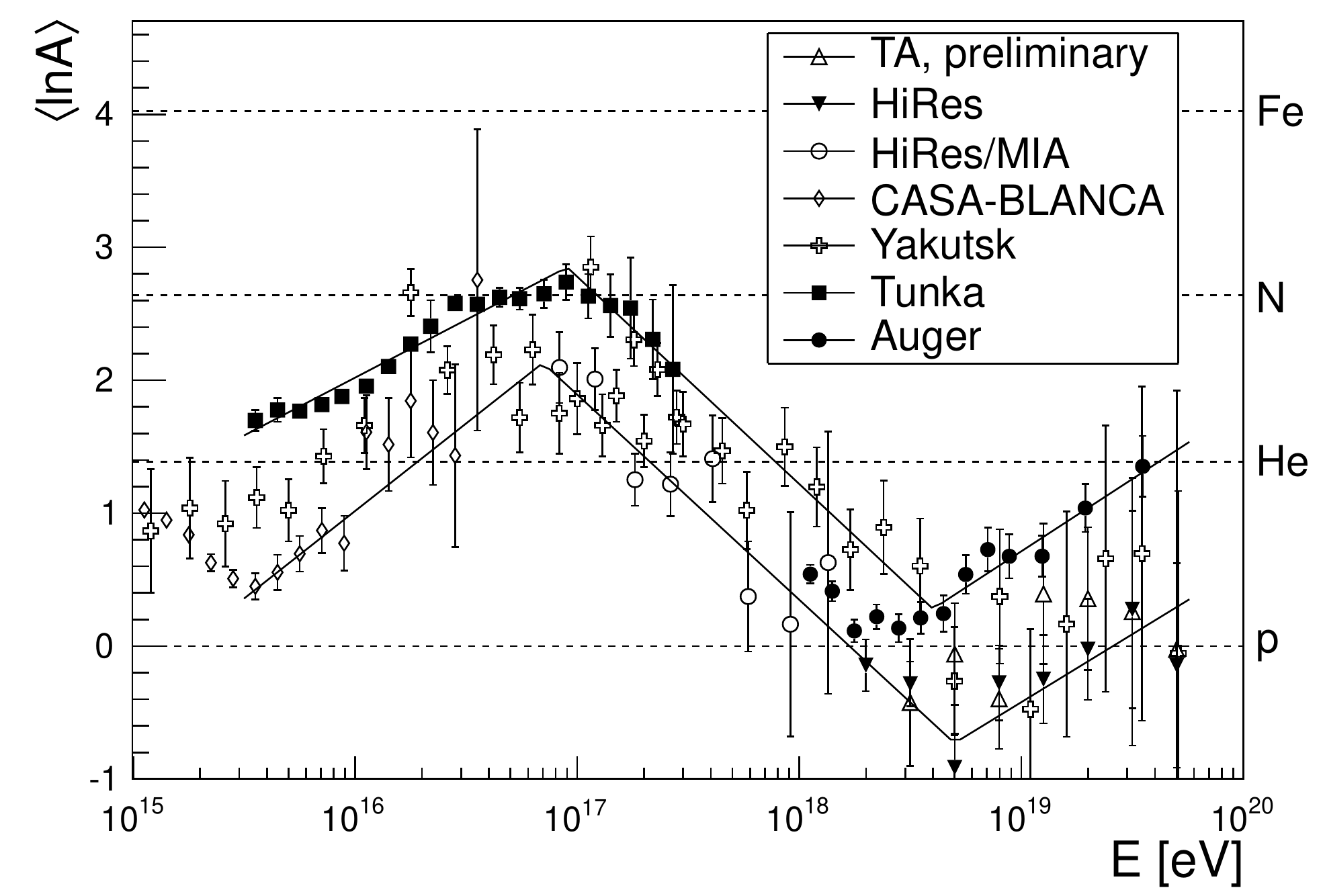}
\label{fig:lnA0}
}
\subfigure[{\scshape QGSJetII}]{
\includegraphics[width=0.45\linewidth]{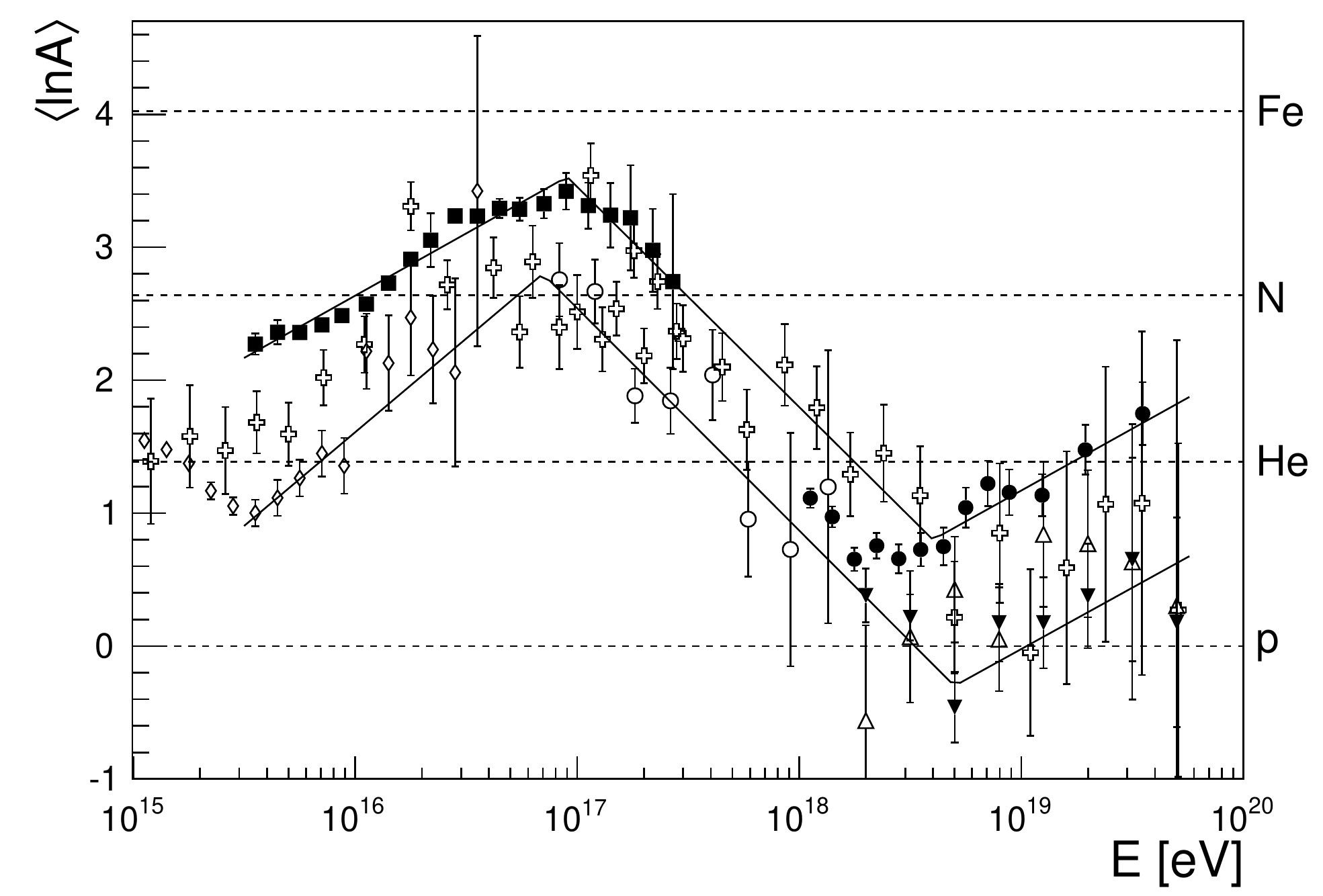}
\label{fig:lnA1}
}
\subfigure[{\scshape Sibyll2.1}]{
\includegraphics[width=0.45\linewidth]{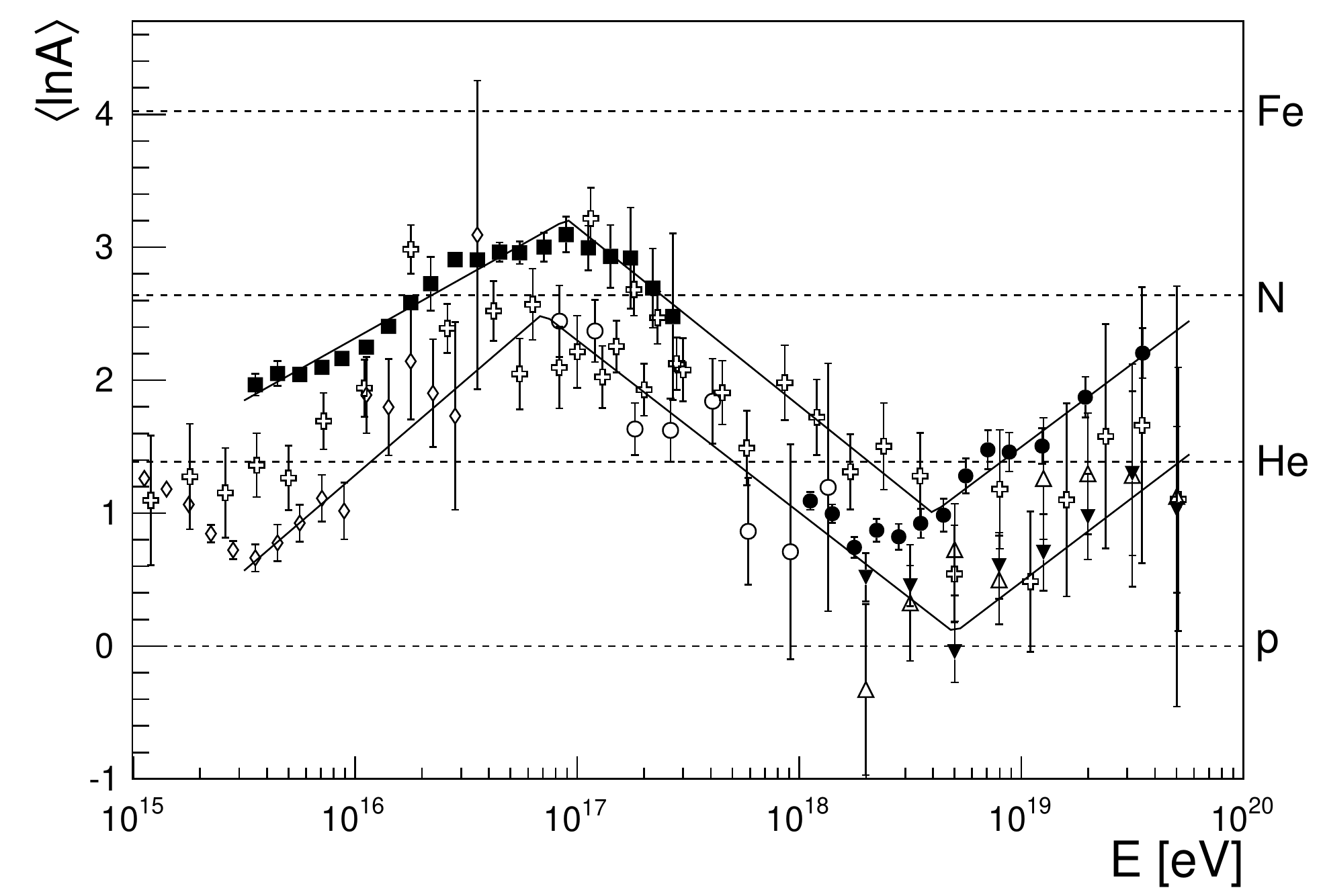}
\label{fig:lnA2}
}
\subfigure[{\scshape Epos1.99}]{
\includegraphics[width=0.45\linewidth]{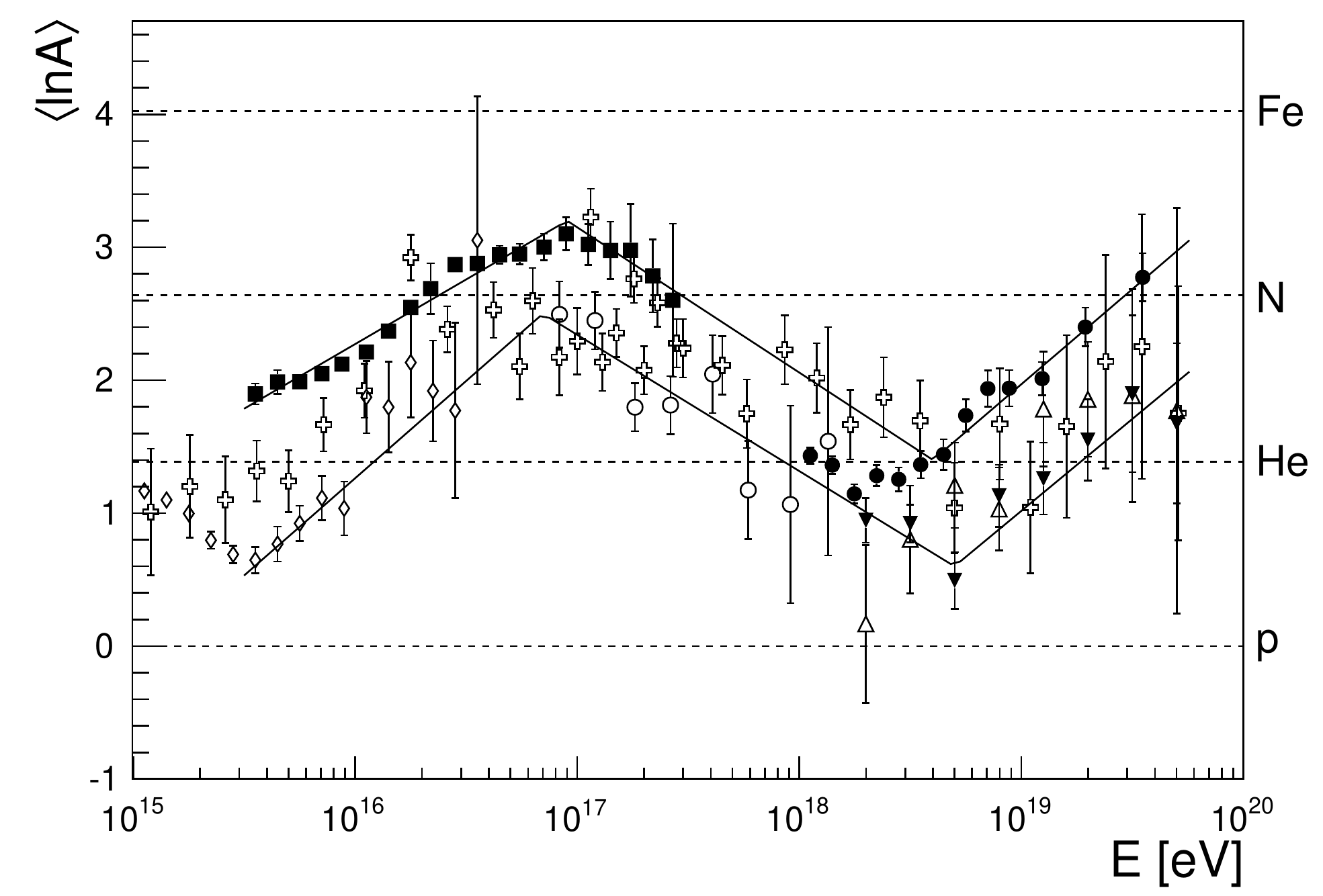}
\label{fig:lnA3}
}
\caption[optical lnA]{Average logarithmic mass of cosmic ray as a function of energy derived
from $\Xmax$ measurements with optical detectors for different hadronic interaction models. Lines are estimates on the experimental systematics, i.e.
upper and lower boundaries of the data presented.}
\label{fig:lnAOptical}
\end{figure*}
As it should have become clear in the previous sections, the
measurement of the mass composition of cosmic rays from air shower
data is a challenging task. There are still considerable systematic
differences between the various air shower measurements that need to
be resolved. A further obstacle arises from the large shower-by-shower
fluctuations of air shower observables which are found to fluctuate at
a level which, for light primaries, may become comparable to the mean
separation of proton and iron primaries. Moreover, and maybe most
importantly, the interpretation of these measurements in terms of the
chemical composition of cosmic rays relies on the validity of models
of hadronic interactions at ultra-high energies.

The best discrimination power in terms of models for the origin of
ultra high energy cosmic rays is of course given by a determination of
the energy dependence of different groups of primary elements. But as
could be seen in e.g.\ Fig.~\ref{fig:FluxPlot}, even when different
experiments use the same hadronic interaction model to interpret their
data, there are large differences in the quoted fluxes of elemental
groups and similar discrepancies can be observed when comparing the
unfolding results using different interaction models. Part of these
difficulties arise due to the large negative correlations between the
unfolded elemental fluxes, which are however not quoted by the
experiments.

We therefore restrict ourselves to a comparison of the average logarithmic
mass of cosmic rays, which we obtain from either
\begin{equation}
  \meanLnA = \sum f_i \ln A_i
\label{eq:lnAFromFractions}
\end{equation}
for experiments that derived elemental groups of mass $A_i$ with flux fractions $f_i$ or from
\begin{equation}
  \meanLnA = \frac{\meanXmaxX{\mathrm{p}} - \meanXmaxX{\mathrm{data}}}
                  {\meanXmaxX{\mathrm{p}} - \meanXmaxX{\mathrm{Fe}}} \ln 56
\label{eq:lnAFromXmax}
\end{equation}
in case of experiments that measured \meanXmax. For the latter, model
uncertainties can be estimated by using different predictions for
$\meanXmaxX{\mathrm{Fe}}$ and $\meanXmaxX{p}$. This
comparison is shown in Fig.~\ref{fig:lnAOptical}, where we also
include the {\scshape QGSJet01} model for completeness, due to the
ubiquity of results from particle detectors that used this superseded
model. Obviously, the systematic differences in \meanXmax discussed in
the last section propagate directly to \meanLnA. To guide the eye and
to be able to compare the results from optical detectors with those of
particle detectors (see below), the upper and lower \meanLnA ranges
are sketched in Fig.~\ref{fig:lnAOptical} by solid lines. As can be
seen, the experimental systematics in \meanXmax translates to an
uncertainty of about $\sigma(\meanLnA)\approx \pm 0.5$. The
composition trends that were already visible in Fig.~\ref{fig:xmax}
can again be observed in \meanLnA: All model interpretations suggest a
gradual increase of the average logarithmic mass of cosmic rays
between \energy{15} and \energy{17} followed by a transition towards a
lighter composition during the next decade. The heaviest composition
with \meanLnA$\approx 3.5$ follows from the Tunka data interpreted
with {\scshape QGSJetII} at around \energy{17}.  The \meanLnA values
of HiRes and TA are compatible with a pure proton composition when
using one of the two {\scshape QGSJet}-flavors. A trend towards a
heavier composition would follow from Auger data for all models and
also for HiRes and TA if interpreted using {\scshape Sibyll} or
{\scshape Epos}.  It is interesting to note that the next version of
{\scshape QGSJetII}~\cite{sergeiICRC11} for which some model
parameters were re-tuned to new data from the LHC will have a similar
$\meanXmax$ as {\scshape Sibyll} and thus the combination of any of
the $\meanXmax$ data with one of the contemporary versions of the
three available interaction models will result in a $\meanLnA$
significantly different from zero at ultra-high energies.

Particle detectors usually do not publish air shower observables but
directly the interpretation in terms of elementary fractions, and in
that case only the differences between models with which the data were
analyzed can be used for a limited estimate of the theoretical
uncertainties. Results that were obtained with out-dated interaction
models like e.g.\ the AGASA measurements~\cite{Hayashida:1995tu} will
be ignored in the following.  Since usually only fractions of
elemental {\itshape groups} are quoted it is not obvious which value
of $\ln A_i$ to assign in Eq.~(\ref{eq:lnAFromFractions}).  To
translate the data from Tibet AS$\gamma$~\cite{Amenomori:2011hu} into
\meanLnA, we assume equal fluxes of protons and helium and assign to
`heavy' fragments $A=32$. However, we note that the chosen procedure
of comparing fluxes from different measurement campaigns with
different event selection and energy calibration may introduce
additional systematic uncertainties particularly in view of the steep
power-law spectra involved, which we can not account for here.  For
KASCADE-Grande~\cite{FuhrmannICRC11}, where the intermediate mass
group is composed of He, C, and Si, we again assume equal fluxes and
take the logarithmic mean of $A\simeq 12$.  For data that were
analyzed in a simple bimodal proton/iron model like~\cite{Dova-04,
  ave-02a} the $\meanLnA$ calculation is technically easy, but it is
difficult to assess the systematic uncertainty arising from this
simplified model.  Data from EAS-TOP are based on electrons and GeV
muons \cite{eastop-04b} as measured in the calorimeter at the surface
as well as on electrons and TeV-muons, the latter measured in MACRO
\cite{eastop-04c}.  Of all the experimental particle detector results
studied here, only Auger published the measured air shower observables
rather than their interpretation.  Since the average muon production
depth and the rise time asymmetries are well correlated with $\Xmax$
we assume that they also depend linearly on $\meanLnA$ and can
therefore use the air shower simulations folded with the detector
response from~\cite{GarciaPinto:2011} and~\cite{GarciaGamez:2011} to
estimate $\meanLnA$ from the equivalent of Eq.~(\ref{eq:lnAFromXmax})
for these variables.

\begin{figure}[ht!]
\centering
\subfigure[{\scshape QGSJet01}]{
\includegraphics[width=0.95\linewidth]{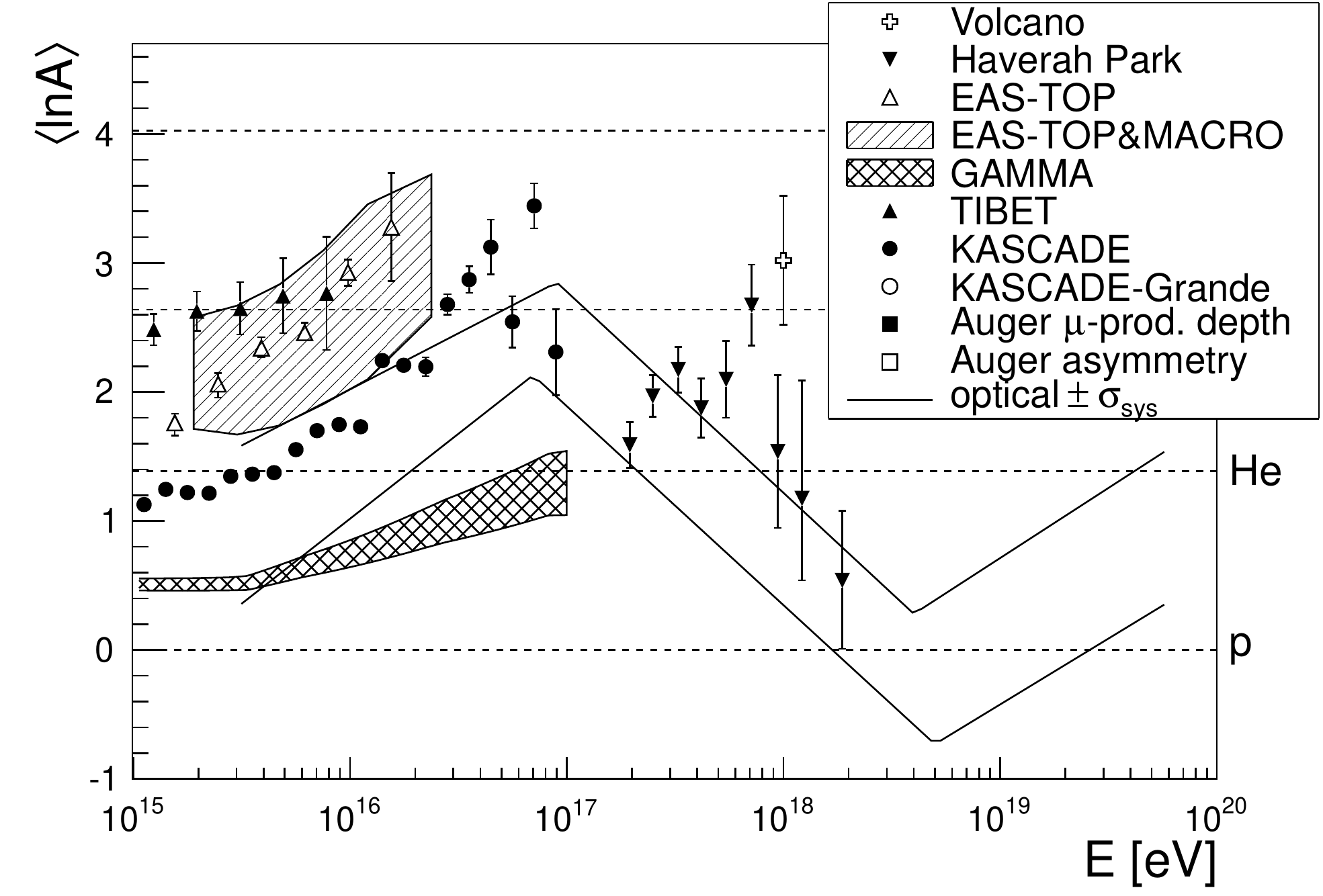}
\label{fig:lnAsurfQG01}
}
\subfigure[{\scshape Sibyll2.1}]{
\includegraphics[width=0.95\linewidth]{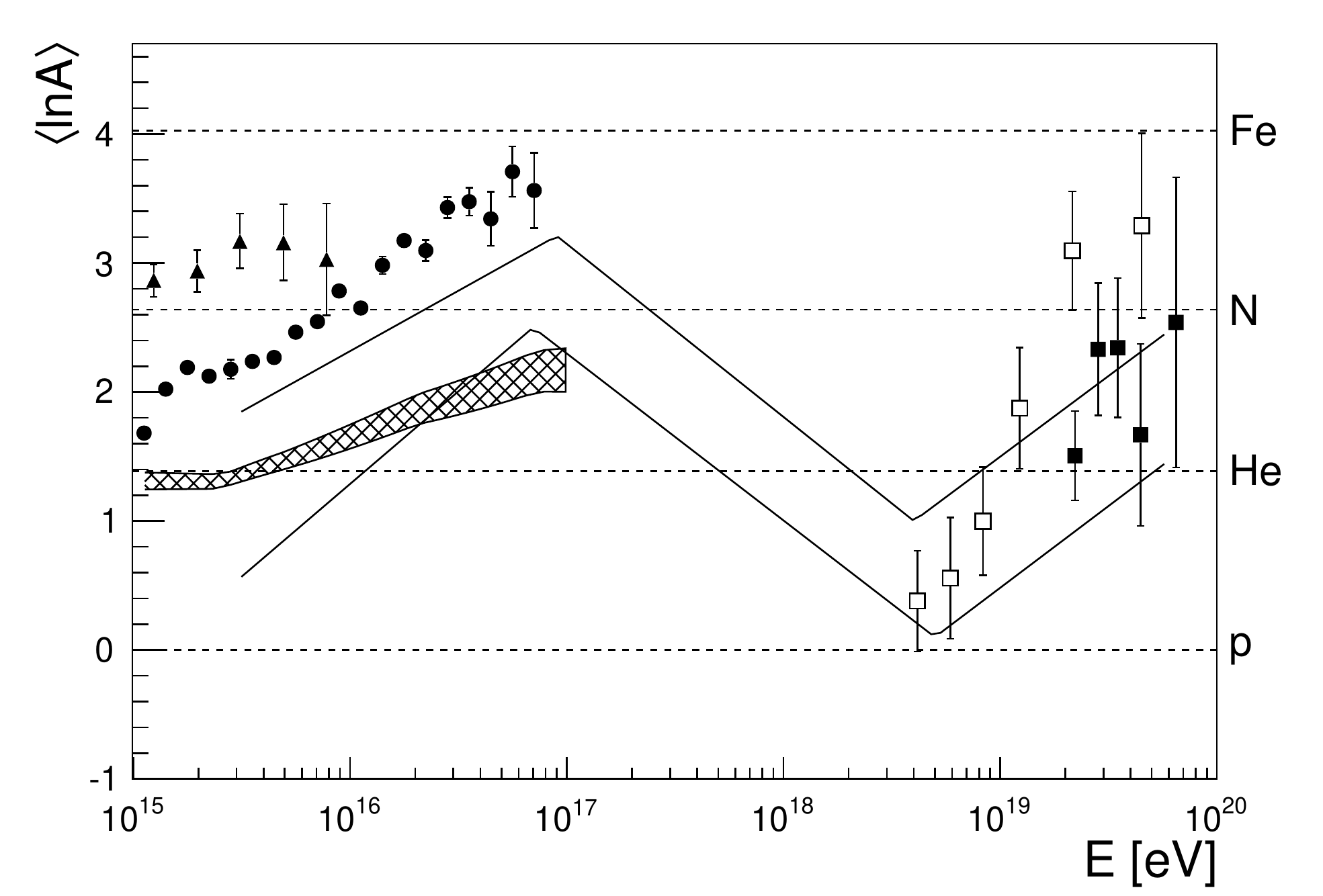}
\label{fig:lnAsurfSib}
}
\subfigure[{\scshape QGSJetII}]{
\includegraphics[width=0.95\linewidth]{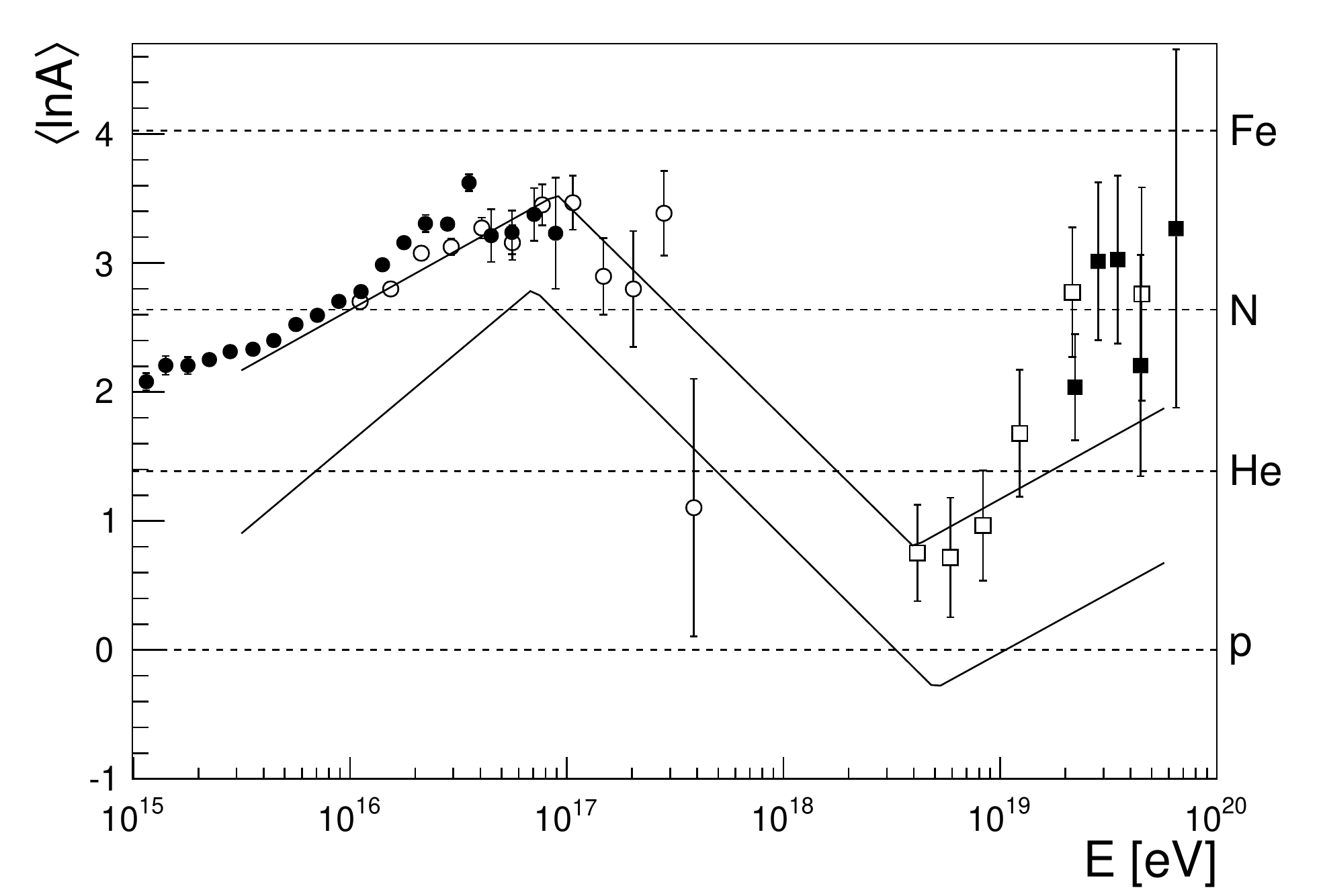}
\label{fig:lnAsurfQGII}
}
\caption[lnA]{Average logarithmic mass of cosmic ray as a function of
  energy derived from particle detector measurements for different
  hadronic interaction models.  Lines are the upper and lower bounds
  from optical measurements (cf. Fig.~\ref{fig:lnAOptical}) for the
  corresponding model.}
\label{fig:lnAsurface}
\end{figure}

The resulting energy evolution of $\meanLnA$ as derived from particle
detector data is displayed in Fig.~\ref{fig:lnAsurface} for different
hadronic interaction models. The upper and lower experimental
boundaries from optical detectors are indicated by the superimposed
lines.  As can be seen, the systematic differences between
experimental results at low energies are considerably larger than in
the case of optical detectors spanning a range of up to
$\Delta\meanLnA\approx \pm 1$.  Nevertheless, all experiments below
\energy{17} report a rise of $\meanLnA$ with energy that could be
reconciled with the $\meanXmax$ results by an appropriate rescaling.
In the energy region toward the ankle, surface detector data are
sparse. The Haverah Park results tend towards a lighter composition at
\energy{18}, though with large statistical uncertainties.  At
ultra-high energies only the surface detector data from Auger are
available for an interpretation with modern hadronic interaction
models. For both simulations, using {\scshape QGSJetII} and {\scshape
  Sibyll2.1}, these data are compatible with an increase of $\meanLnA$
above \energy{19}.

%% file: neutral.tex
\begin{figure}[b!]
  \centering
  \includegraphics[width=0.75\linewidth]{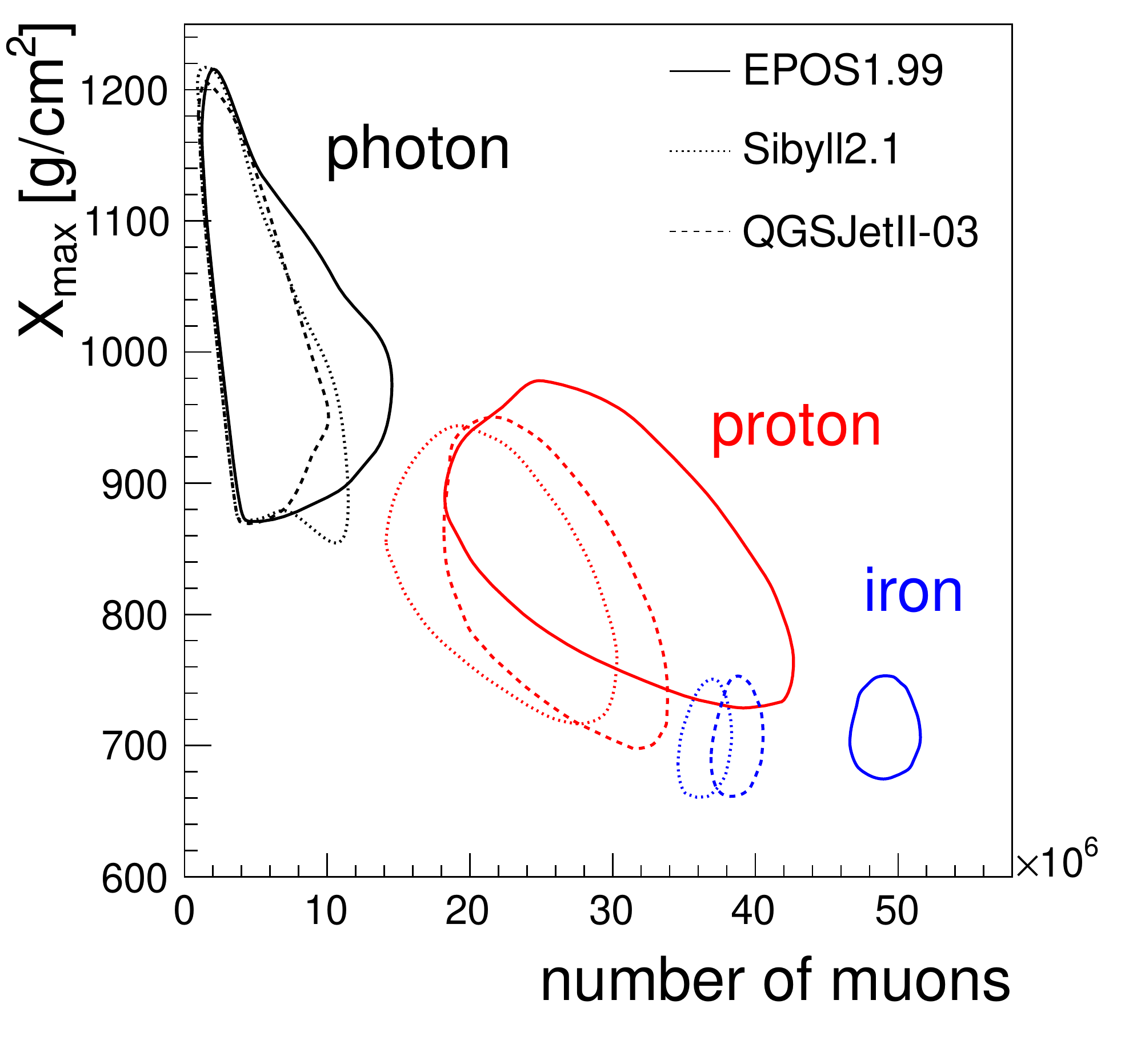}
  \caption[xmaxVsNmu]{Simulation of hybrid air shower observation:
    Shower maximum vs.\ number of muons at ground level (60$^\circ$
    shower at 800\,\gcm) for primary energies of \energy{19}. Contour
    lines illustrate the regions which include 90\% of the showers. }
  \label{fig:xmaxVsNmu}
\end{figure}
\subsection{Photons}

Air showers induced by primary photons develop an almost pure
electromagnetic cascade and differ from hadron induced ones by their
low number of muons and their deep \Xmax. This is illustrated in Fig.\
\ref{fig:xmaxVsNmu} for different primaries at \energy{19}. Since
mostly electromagnetic processes are involved in the shower
development, the predictions are more reliable and do not suffer from
uncertainties in the hadronic interaction models. The Pierre Auger
Observatory has the advantage of being a hybrid observatory in which
the position of the shower maximum is measured by fluorescence
telescopes with simultaneous access to the muon number from the
surface detector stations. In case muons cannot be identified, such as
in pure scintillator arrays located at the surface, the particle
densities at large distances from the shower core combined with
measurements of the energy by e.g.\ fluorescence telescopes can still
provide good discrimination power.

\begin{figure}[t!]
  \includegraphics[width=\linewidth]{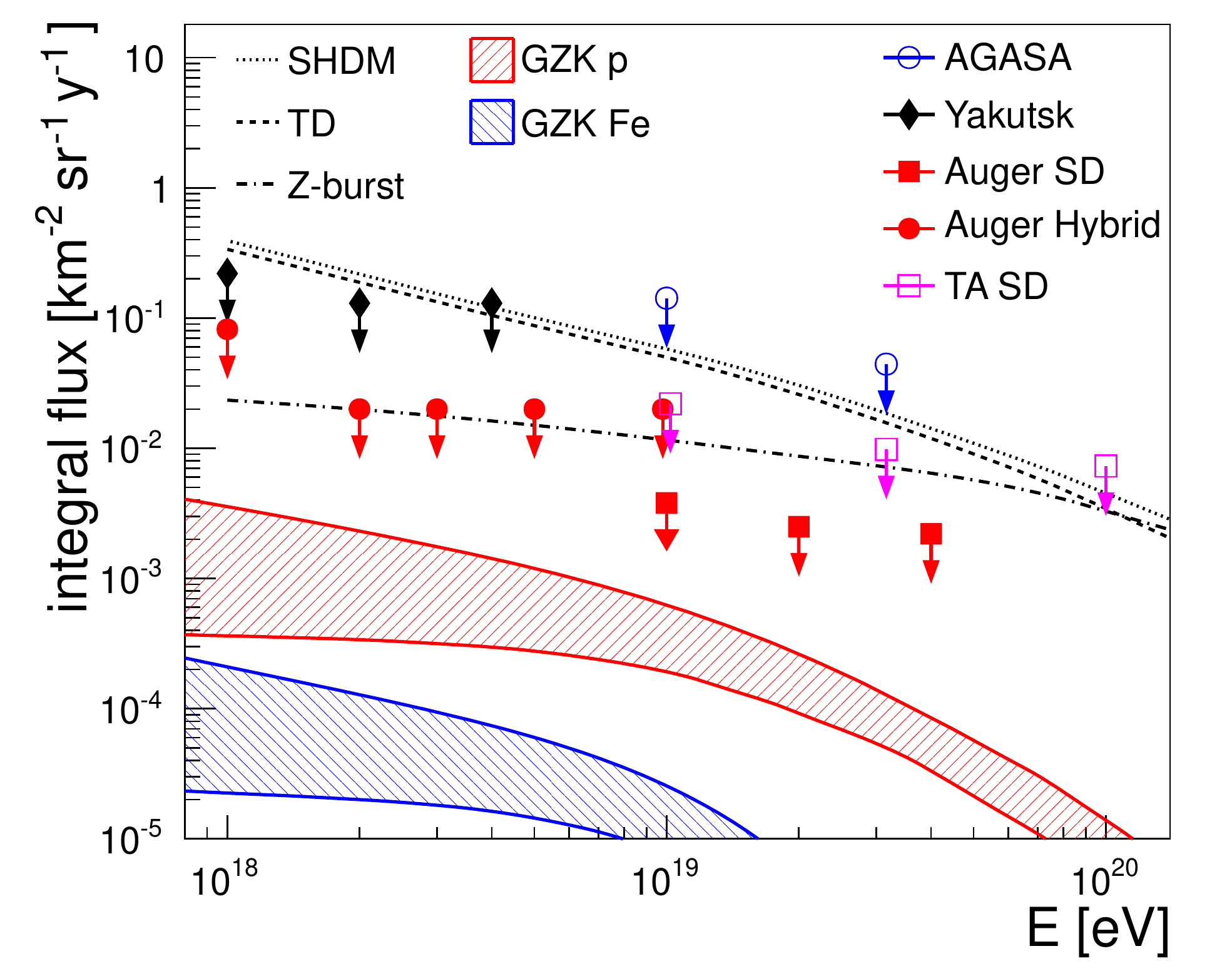}
  \caption[photon]{Integral photon flux limits at 95\% C.L.\ from
    AGASA~\cite{Shinozaki:2002ve}, Yakutsk~\cite{Glushkov:2009tn},
    Auger~\cite{Aglietta:2007yx, augerPhotonICRC11} and
    TA~\cite{taPhotonICRC11} compared to flux predicitons for
    GZK-photons~\cite{Sarkar-11}, top-down scenarios of
    super-heavy dark matter (SHDM)~\cite{Ellis:2005jc} and
    topological defect (TD) models, and 
  $Z$-bursts~\cite{Gelmini:2008fs}.}
  \label{fig:photon}
\end{figure}

Early searches for UHE photons, summarized e.g.\ in
Ref.~\cite{Risse-07}, were motivated by particle physics
models of UHECR origin. In these so-called top-down models (see
e.g.~\cite{bhattacharjee99} for an early review) the highest energy
cosmic rays are decay products of super-heavy relic particles or
topological defects (TD) left over from the inflationary epoch
and which are locally clustered in the galactic halo as cold
dark matter. Their decay yields a relatively high flux of UHE
photons and neutrinos reaching up to 90\,\% of the CR flux.
The Z-Burst model~\cite{weiler99,Fodor:2001qy} involves
resonant interactions of very energetic neutrinos with the relic
neutrino background. These models are compared to experimental data
in Fig.\ \ref{fig:photon}. The integral photon flux limits from
AGASA~\cite{Shinozaki:2002ve}, Yakutsk~\cite{Glushkov:2009tn}, and
TA~\cite{taPhotonICRC11} are based on muon numbers and curvature
of the shower front.
The hybrid data of the
Pierre Auger Observatory, employing \Xmax and various SD-observables,
provide the best present limits up to \energy{19} while the Surface
Detector data alone extend the best present limits up to $5\times
10^{19}$~eV due to the much higher statistics~\cite{Aglietta:2007yx,
  augerPhotonICRC11} available in this data set.

The results demonstrate that particle physics motivated top-down
models are strongly disfavored giving support to an astrophysical
origin of UHECR.

The two bands below the experimental data depict photon fluxes
expected from interactions of UHECR with photon background fields,
most prominently the Cosmic Microwave Background (CMB)
radiation. Both, proton and iron primaries have been considered in
these calculations. A power law energy spectrum is assumed with index
$\gamma = -2$ and a maximum CR energy at the source of
\energy{21}. The fluxes are normalized to the number of events
measured by Auger \cite{Abraham-10-hybspec} for energies $E > 
10^{18.4}$\,eV.
The upper and lower bounds for each primary result from
more or less favorable radio background fields and different cosmological
source evolutions~\cite{Sarkar-11}. According to these simulations,
chances of observing GZK-photons appear well in reach if
the source spectrum at the highest energies is dominated by light
primaries. Moreover, further improvements, besides increasing statistics
of data, may be expected from multivariate analyses of fluorescence and
surface detector data. Also, further upward and downwards
modifications of the predicted photon flux may result
from  different energy spectra and/or different maximum energies
of the nearby sources~\cite{Sarkar-11,Hooper:2011kr}.

\begin{figure}[t!]
  \includegraphics[width=\linewidth]{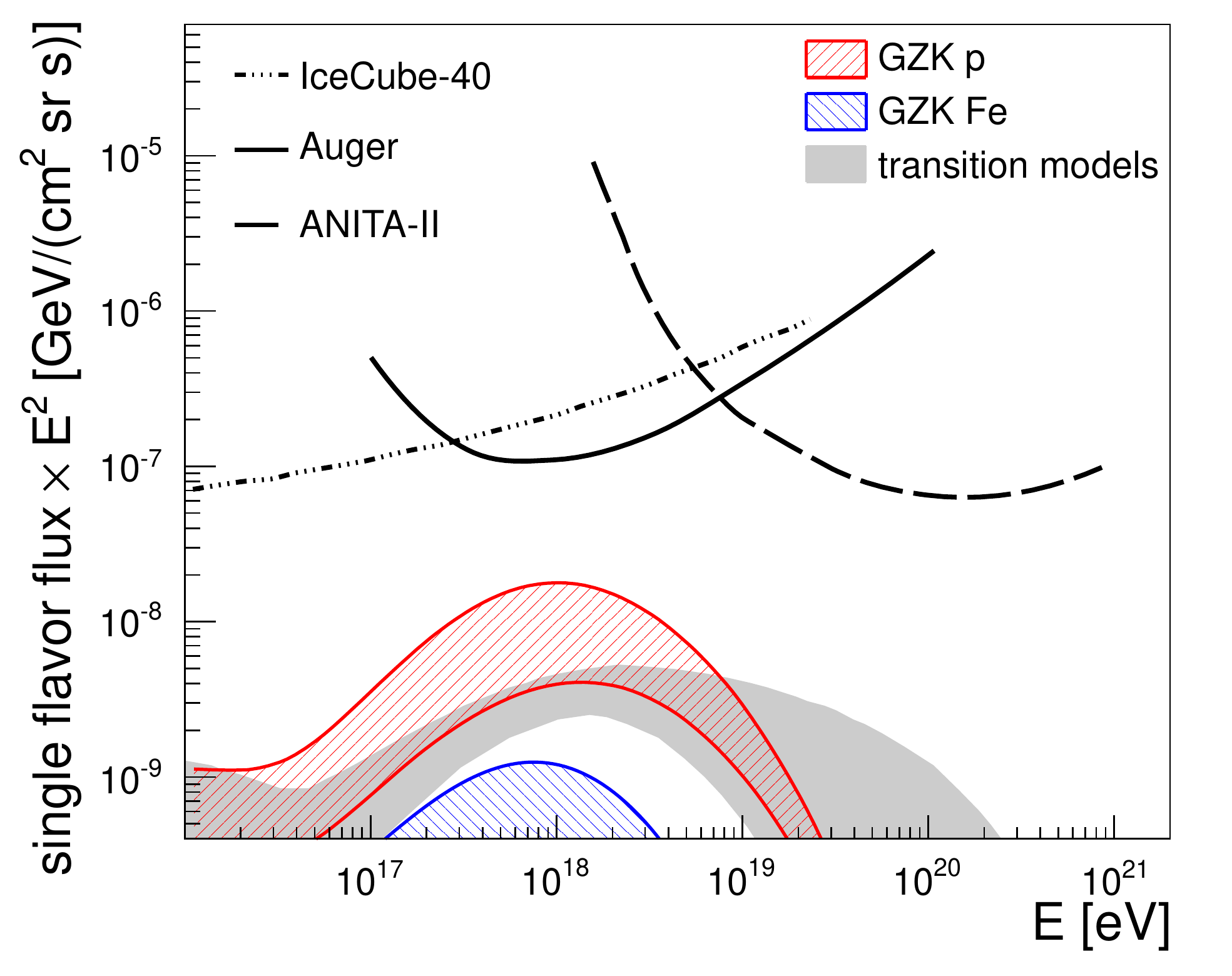}
  \caption[neutrino]{Compilation of 90\,\%CL single-flavor upper
    limits for diffuse neutrino fluxes assuming a proportion of
    flavors of 1:1:1 due to neutrino oscillations. Data are from
    IceCube~\cite{Abbasi:2011ul}, Auger~\cite{icrc0682} and
    ANITA~\cite{Gorham-10}. The shaded area corresponds to expected
    GZK neutrino fluxes computed under different assumptions of source
    evolution scenarios~\cite{Sarkar-11} with
    power-law energy spectra of $\gamma=-2.0$ and $E_{\rm max}^Z = Z
    \cdot 10^{20}$~eV. The grey band depicts
    different transition models and source evolutions adapted from
    Ref.~\cite{Kotera:2010hx}
    (see text for details).\label{fig:neutrino}}
\end{figure}
\subsection{Neutrinos}

Possible neutrino primaries may be the easiest to identify in EAS
experiments because of the many orders of magnitude difference
between the electroweak and hadronic cross sections. With a
neutrino-nucleon cross section of about $\sigma_{\nu N} \simeq 10$~nb
at 1\,\EeV, neutrinos may interact at any point in the atmosphere or
even in the rock of the Earth. Thus, showers starting very deep in the
atmosphere are likely to be initiated by a primary
neutrino. Identification of such showers is easiest in horizontal
direction ($\theta > 85^\circ$) where the atmospheric thickness is
effectively 30 times that of the vertical direction. Thus, horizontal
showers containing still an appreciable electromagnetic component
(so-called `young showers') can only be caused by primary
neutrinos. Larger experimental apertures than for near-horizontal showers
induced by so-called `down-going' neutrinos
are reached for `up-going' tau neutrinos skimming the
Earth at zenith angles between $90^\circ$ and $96^\circ$. After a
possible $\nu_\tau$-interaction, the produced $\tau$ will be able to
escape the Earth and decay in the atmosphere or close to ground
producing an upward-going EAS.

Such signatures have been searched for in ground based EAS
observatories such as Auger~\cite{Abraham-09a} and
HiRes~\cite{Abbasi:2008hr} reaching highest sensitivities around
\energy{18}, thus matching very well the peak in the differential
$E_\nu^2\text{d}j_\nu/\text{d}E_\nu$ fluxes of cosmogenic neutrinos.
At even higher energies, radio based observations such as
ANITA~\cite{Gorham-10} start to take over, while IceCube provides
unprecedented sensitivities towards lower energies. The
three experimental approaches complement each other very effectively
with each of them having improved their sensitivities by orders of
magnitudes in the last decade (cf.\ Fig.\ \ref{fig:neutrino}).

Similarly as for UHE photons, these searches did not detect any
neutrino fluxes yet but provided upper limits. Fig.~\ref{fig:neutrino}
depicts some exemplary reference predictions again for sources
emitting proton and iron primaries~\cite{Sarkar-11,Kotera:2010hx}. Due
to their low cross-section, neutrinos --- other than photons ---
suffer only adiabatic energy loss, so that they can arrive at Earth
from cosmological distances. As a result, a large level of uncertainty
in the flux prediction results from the unknown cosmological evolution
of source luminosity, as has been realized already
in~\cite{Engel:2001iy}. The upper dashed bands shown in
Fig.~\ref{fig:neutrino} are obtained for an FR-II type distribution
while the lower one results for a star formation like evolution. As
for the photon predictions, power law source distributions with an
injection index of $\gamma = -2$ have been assumed as a reference with
$E_{\rm max}^Z = Z \cdot 10^{20}$~\eV. A fit to the Auger CR energy
spectrum performed e.g.\ in~\cite{Sarkar-11,Decerprit:2011bt} yields
somewhat steeper source spectra but higher maximum energies of the
sources with essentially compensating effects with regard to the
neutrino fluxes.  The grey band, adapted from Ref.\
\cite{Kotera:2010hx}, depicts neutrino predictions covering different
transition models (dip-model at top and ankle model at bottom) and
source evolutions.

Atmospheric neutrinos in the \TeV to \PeV range are primarily of
interest, because they constitute the background in neutrino
telescopes such as IceCube~\cite{Abbasi:2011ul} or
Antares~\cite{Agerona:2011nsa}. The cosmic rays producing this
`calibrated' beam of neutrinos most dominantly originates from
energies above the knee. Thus, an interesting question to ask is the
level of uncertainty in the neutrino flux calculations that arises
from the poorly known composition (see e.g.~\cite{Illana:2011km}).
Expressed differently, one may ask whether a comparison of the
measured atmospheric neutrino spectra with flux calculations based on
cosmic ray energy spectra could provide independent information about
the composition at the knee and above.  A quantitative study of this
question has been presented very recently in
Ref.~\cite{Bindig-ICRC11}. It used the two different sets of energy
spectra shown in Fig.~\ref{fig:FluxPlot} from KASCADE
\cite{KASCADE-05} combined with direct measurements of cosmic rays at
lower energies as input to a full Monte Carlo simulation. Based on
this study, the authors concluded that uncertainties in the
all-particle cosmic ray flux appear more important for the neutrino
fluxes than the actual uncertainties of the composition arising from
the unfolding based on {\scshape QGSJet} versus {\scshape Sibyll}
interaction models.

%% file: conclusions.tex
\begin{figure*}[t!]
\centering
\includegraphics[width=0.8\linewidth]{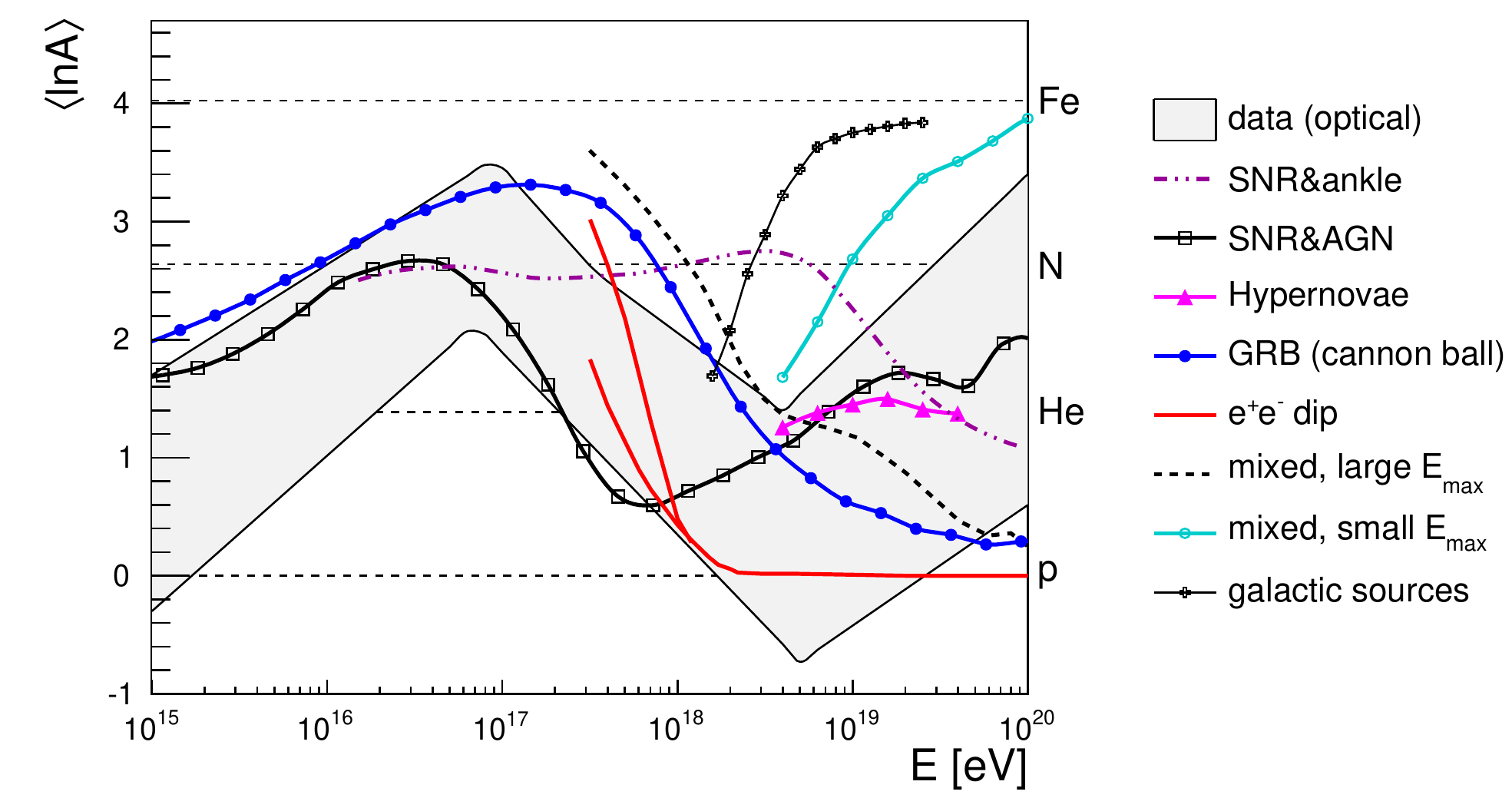}
\caption[lnA and models]{Compilation of \meanLnA-values from
recent astrophysical models compared to the uncertainty range of experimental data (optical detectors, see text).
Models include the purely galactic scenario of~\cite{Calvez:2010uh},
a mixed extra-galactic composition
with a large~\cite{Allard:2005cx} or small~\cite{Allard-11}
maximum energy of the sources, the ankle and dip model
\cite{Berezinsky:2002nc}, the cannonball model \cite{dar-06},
a hypernova model~\cite{Liu:2011us}, and the SNR-AGN model of
\cite{Berezhko:2009kq}.\label{fig:lnaAndModels}}
\end{figure*}

In 1983, Linsley counted the number of contributions submitted to
the 18th ICRC that were related to the composition of cosmic rays and
concluded~\cite{LinsleyXmax1983}
\begin{quote}
  \itshape Assuming that people wouldn't spend effort on this problem
  unless they believe it can eventually be solved, I think the numbers
  are encouraging.
\end{quote}
Today, almost 30 years later, the quest for the understanding of the
primary composition of cosmic rays continues and, although the problem
is not fully solved yet, there has been considerable progress in at least
our qualitative understanding of the primary cosmic ray composition
thanks to the wealth of new data collected by both cosmic ray
observatories and particle physics experiments.

At the energies between \energy{15} and \energy{17} all experiments
observe energy dependent changes in the shower development that are
compatible with increasing average mass of cosmic rays.
Moreover, unfolded spectra of mass groups in this energy range suggest
that the change of composition is attributable to a consecutive cut-off in the
flux of the individual mass components starting with protons at a few
\energy{15} up to iron at around \energy{17}
(cf.\ Figs.~\ref{fig:FluxPlot} and \ref{fig:fluxKA-Grande}). These
composition measurements are thus compatible with an interpretation of
the knee in the particle flux due to a rigidity dependent leakage of
cosmic rays from the galaxy and/or a rigidity dependent maximum energy
of galactic sources.  At the same time, a particle physics
interpretation of the knee is disfavored, since the interaction models
used for the interpretation of cosmic ray data were found to bracket
particle production measurements from the LHC~\cite{dEnterria:2011kw}
at 7~\TeV center of mass energy, corresponding to a primary cosmic ray
energy of \energy{16.4}.

Towards the energy region of the ankle, air shower measurements
indicate a decrease of the average mass of cosmic
rays. Both the shower-to-shower fluctuations and the average shower
maximum are compatible with a predominantly light composition at a few
\energy{18}.  Estimates of the proton-air cross
section~\cite{Knurenko:1999cr,Belov:2006mb,AugerXsecICRC11} at this
energy agree well with the extrapolations used in hadronic interaction
models, therefore also here a drastic change of the interpretation of
the measurements in terms of cosmic ray composition due to
uncertainties in air shower simulations seems unlikely.

At the highest energies, above \energy{19}, the experimental uncertainties
are still too large to draw firm conclusions from the data.
The measurements from Auger of $\meanXmax$, $\sigmaXmax$, muon
production depth and rise-time asymmetries may be interpreted as
a transition to a heavier composition that may be caused by a
Peters-cycle in extra-galactic sources similar to what has been
observed at around the knee. However, the
$\meanXmax$-measurements from HiRes, TA and Yakutsk
indicate a systematically lighter composition
at these energies but with an elongation rate compatible with Auger
(cf.\ Tab.~\ref{tab:XmaxFit}).

These experimental differences together with the uncertainties of
hadronic interactions make an astrophysical interpretation of
the \meanLnA estimates at the highest energies difficult.
In Fig.~\ref{fig:lnaAndModels}
we present a compilation of astrophysical models for the composition of
cosmic rays together with the \meanLnA inferred from
air shower measurements.
Here, we show only results from optical measurements, since
these are more abundant over the full energy range and ---
judging from the differences of results from surface detectors at low
energies in Fig.~\ref{fig:lnAsurfQG01} --- are also less affected by
experimental systematics. The gray band is the maximum and minimum of
the envelopes from Fig.~\ref{fig:lnAOptical}, i.e.\ the upper and lower
level of experimental {\itshape and} model differences. The curves
represent predictions by recent models about the origin of cosmic
rays with a focus on models at ultra-high energies (see~\cite{Horandel:2005bb}
and references therein for a comprehensive list of models of cosmic rays around the knee).
A fairly good description of \meanLnA over the
entire energy range is given by the two-component model of~\cite{Berezhko:2009kq}
in which it is assumed that CRs up to
\energy{17} are produced in galactic supernova remnants while the dominant component
at higher energies is of extragalactic origin produced at the shock
created by the expanding cocoons around active galactic nuclei.
The dashed violet curve shows the classical ankle model and the two red
lines the so-called dip-model~\cite{Berezinsky:2002nc} (actual calculation used are
from ~\cite{Allard:2005cx}). Other than
the ankle model, the dip model assumes that the transition from
galactic to extragalactic models occurs below the ankle and that the
ankle is caused by $e^+ e^-$-interactions of protons in the CMB rather
than by the onset of the extragalactic cosmic ray component. To
make this Bethe-Heitler process work, the composition above \energy{18}
must be dominated by protons.
The blue curve represents the generalized cannonball model~\cite{dar-06}
in which cosmic rays are described as being ions of the interstellar
medium that encountered cannonballs --- highly relativistic bipolar
jets of plasmoids originating from supernova explosions and GRBs ---
and were magnetically kicked up to higher energies.
In Ref.~\cite{Liu:2011us} (shown as the full magenta
line with triangles) it has been suggested that hypernova remnants,
with a substantial amount of energy in semi-relativistic ejecta, can
accelerate intermediate mass or heavy nuclei to ultra-high energies.
A heavy composition at the highest energies is obtained in a model
where UHECRs were produced by gamma-ray bursts or rare types of supernova
explosions that took place in the Milky Way in the past~\cite{Calvez:2010uh}.
Finally, the models labels `mixed' with large and small $E_{\rm max}$,
are taken from Refs.~\cite{Allard:2005cx,Allard-11}. Here, a source
composition similar to galactic CRs is assumed with the ankle marking
again the transition from galactic to extragalactic CRs. In the first
realization of the model it is assumed that the maximum energy follows
$E_{\rm max}^{\rm high} = Z\cdot10^{20.5}$\,\eV with an injection
index at the source of $\gamma=-2.0$~\cite{Allard:2005cx} whereas in
the latter $E_{\rm max} = Z\cdot4\times10^{18}$\,\eV and
$\gamma=-1.6$~\cite{Allard-11}.

While data and models agree reasonably well up to the ankle energy,
there is almost the full range of masses covered by models at energies
above \energy{19}. This may be understood from the large uncertainties of
experimental data points in this energy range. At present neither a
composition dominated by light primaries nor by heavy primaries can
safely be ruled out by the data.

As discussed in this paper, a large part of these uncertainties can be
attributed to the uncertainties of hadronic interaction models, but
also systematic differences between the experimental data itself
contribute significantly to the large width of the \meanLnA-range at
the highest energies. Fortunately, hadronic interaction models will
improve considerably in the near future due to the new particle
production data from LHC.  At the same time further improvements of
the quality of air shower data are expected because of better
instruments and the prospects of a combined analysis of the data of
surface and optical detectors with either existing hybrid detectors or
enhancements such as e.g.\ shielded muon detectors in
Auger~\cite{amigaICRC11}.

Additional constraints about the charge of cosmic rays may be
acquired from the anisotropy (or lack thereof) of the arrival
directions of cosmic rays. The
Pierre Auger Collaboration has reported directional correlations of
the most energetic particles ($E > 5.5\times10^{19}$\,eV) with the
positions of nearby AGN~\cite{Abraham-07e,Abraham-08a,Abreu-10}. Since 
the
turbulent component of the galactic magnetic field randomizes
the arrival directions of particles with small rigidities (i.e.\ large charge
and therefore mass), such a correlation seems to suggest a dominance of
light particles at these energies, but it may be still possible
that a large scale anisotropy, e.g.\ from the super-galactic plane, remains
anisotropic even after the passage of heavy particles through the
turbulent magnetic field~\cite{Giacinti:2011uj}.

A next generation giant observatory
should be able to collect large statistics for both, charged particle astronomy and
composition analyses above the GZK threshold.
If cosmic rays are extragalactic and their flux at earth is not
dominated by very close sources, then the composition analysis at
GZK-energies will be facilitated by the suppression of any
intermediate nuclei between proton and iron at earth due to the
different energy loss lengths of
nuclei~\cite{Allard:2005cx}. In such a simple situation with a 'guaranteed'
composition of protons and iron only  it should be possible to simultanously
determine the properties of cosmic rays {\itshape and} hadronic interactions
at energies above $\sim 6\times$\energy{19}.

So without any doubt, measurements of the cosmic ray composition
up to \energy{20} remain one of the most important tasks for the future and the key
to the understanding of the origin of the most energetic particles in the
Universe.